\documentclass[prx,aps,twocolumn,nopacs,superscriptaddress,longbibliography,nofootinbib]{revtex4-1}

\usepackage{amsmath}  \usepackage{amssymb}  \usepackage{amsfonts}  \usepackage{bm}  \usepackage{bbm}   \usepackage{bbold}   \usepackage{braket}  \usepackage{color}  \usepackage{comment}  \usepackage{dcolumn}  \usepackage{enumerate}  \usepackage{epsfig}  \usepackage{gensymb}  \usepackage{graphicx}  \usepackage{indentfirst}  \usepackage{lmodern}  \usepackage{mathrsfs}  \usepackage{mathtools}  \usepackage{psfrag}  \usepackage{pst-all}  \usepackage{soul}  \usepackage{xcolor}
\usepackage{float} 
\usepackage[colorlinks,linkcolor=blue,citecolor=blue,urlcolor=blue,hyperindex,driverfallback=dvipdfm]{hyperref}  \usepackage[T1]{fontenc} 

\def\ii{{\rm i}}  \def\ee{{\rm e}}

        \def\Eb{{\bf E}}                  \def\jb{{\bf j}}                  \def\Rb{{\bf R}}  \def\rb{{\bf r}}      \def\vb{{\bf v}} 
\def\xx{\hat{\bf x}}  \def\yy{\hat{\bf y}}  \def\zz{\hat{\bf z}}            
\def\kpar{k_\parallel}   
\def\EF{{E_{F}}}     
      \def\wp{\omega_{p}}   
        
\def\EE{\mathcal{E}}

\begin{document}
\title{Generation of entangled waveguided photon pairs by free electrons}

\author{Theis~P.~Rasmussen}
\altaffiliation{These authors contributed equally to this work}
\affiliation{POLIMA---Center for Polariton-driven Light--Matter Interactions, University of Southern Denmark, Campusvej 55, DK-5230 Odense M, Denmark}

\author{\'Alvaro~Rodr\'{\i}guez~Echarri}
\altaffiliation{These authors contributed equally to this work}
\affiliation{ICFO-Institut de Ciencies Fotoniques, The Barcelona Institute of Science and Technology, 08860 Castelldefels (Barcelona), Spain}
\affiliation{Max-Born-Institut, 12489 Berlin, Germany}

\author{Joel~D.~Cox}
\affiliation{POLIMA---Center for Polariton-driven Light--Matter Interactions, University of Southern Denmark, Campusvej 55, DK-5230 Odense M, Denmark}
\affiliation{Danish Institute for Advanced Study, University of Southern Denmark, Campusvej 55, DK-5230 Odense M, Denmark}

\author{F.~Javier~Garc\'{\i}a~de~Abajo}
\email[Corresponding author: ]{javier.garciadeabajo@nanophotonics.es}
\affiliation{ICFO-Institut de Ciencies Fotoniques, The Barcelona Institute of Science and Technology, 08860 Castelldefels (Barcelona), Spain}
\affiliation{ICREA-Instituci\'o Catalana de Recerca i Estudis Avan\c{c}ats, Passeig Llu\'{\i}s Companys 23, 08010 Barcelona, Spain}

\begin{abstract}
Entangled photon pairs are a key resource in future quantum-optical communication and information technologies. While high-power laser light propagating in bulk nonlinear optical crystals is conventionally used to generate entangled photons that are routed into optical configurations, such schemes suffer from low efficiency due to the weak intrinsic nonlinear optical response of known materials and losses associated with photon in- and out-coupling. Here, we propose a scheme to generate entangled polariton pairs directly within optical waveguides using free electrons, whereby the measured energy loss of undeflected electrons heralds the production of counter-propagating polaritons pairs that are entangled in energy and direction of emission. As a paradigmatic example, we study the excitation of plasmon polaritons in metal strip waveguides that, within specific frequency regimes, strongly enhance light-matter interactions that lead to two-plasmon generation in comparison to the probability of single-plasmon excitation. We demonstrate that, under appropriate conditions, an electron energy loss detected in an optimal frequency range can reliably signal the generation of a plasmon pair entangled in energy and momentum. Our proposed scheme can be directly applied to other types of optical waveguides for \textit{in situ} generation of entangled photon pairs in quantum-optics applications.
\end{abstract}
\maketitle

\section*{Introduction}

Entanglement---the nonlocal quantum correlation between particles or quanta---constitutes a key resource in emerging quantum information and communication technologies, particularly in the areas of computation \cite{BD00}, cryptography \cite{JSW00}, and teleportation \cite{MHS12}. The fragility of quantum states unfortunately renders entangled particles extremely sensitive to decoherence introduced by environmental factors. Photons, which interact weakly with their surrounding environment and propagate at the ultimate speed of light, are widely used as carriers of quantum information that is typically encoded in their polarization or momentum states \cite{MVW01,FLP12,KGK23}. Entangled photon pairs are routinely generated through spontaneous parametric down-conversion, a second-order nonlinear process that conserves both the photon polarization and momentum \cite{KMW95,KWW99,AB00,B20_2}. The low efficiency of nonlinear optical processes however necessitates the propagation of intense phase-matched laser light in bulk crystals to generate a significant response, thus circumscribing the generation of down-converted photons to macroscopic material platforms from which entangled pairs are collected and routed in cumbersome optical setups.

Numerous strategies have been proposed to circumvent the undesired decoherence and losses associated with photon in- and out-coupling, for example, through the efficient \textit{in situ} generation of entangled photon pairs within an optical waveguide via nonlinear optical processes \cite{ISM04,YLS08,PSG09,HSS15,LHK15,GZS17}. In particular, the counter-propagating guided photon pairs produced by spontaneous parametric down-conversion under plane wave illumination impinging normally on an optical waveguide are entangled through momentum conservation \cite{paper385,SBG22}. In addition, the possibility of generating entangled photons directly in nanophotonic architectures presents appealing prospects for developing integrated quantum-optical devices \cite{TMK12}. In this context, polaritons---hybrid light-matter excitations involving polarization charges---offer the means to confine and manipulate light on nanometer lengthscales, well below the optical diffraction limit. These excitations can be guided by engineering low-dimensional materials to form polaritonic waveguides with enhanced dispersion \cite{FS15,RGX21}. Notably, the extreme light concentration associated with plasmon polaritons supported in ultrathin or two-dimensional (2D) materials can substantially boost the intrinsic optical nonlinearity of their host, potentially enabling nonlinear light-matter interactions on the few-photon level \cite{paper226,SBG22,CJR23}. Although polariton-based schemes favor mode volume at the expense of quality factor compared to dielectric waveguides, both platforms rely on relatively intense optical fields to trigger the generation of entangled photons.

As an alternative to far-field light sources, energetic electrons supply broadband evanescent electromagnetic fields that can be focused with high spatial precision to explore light-matter interactions at the nanoscale \cite{paper149}. The light emitted by bombarding a photonic structure with electrons is measured in cathodoluminescence (CL) spectroscopy to spatially and spectrally map optically active modes \cite{paper251}, while nonradiative excitations in matter can be probed in electron energy-loss spectroscopy (EELS) by measuring the changes produced in the energy and momentum distributions of scattered electrons \cite{PSV1975,E96}. The ability of free electrons to generate heralded single photons with unity-order efficiency was predicted by introducing the concept of phase-matched propagation in a scheme where free electrons pass tangentially to a curved waveguide \cite{paper180}, a possibility that has been realized in recent experiments \cite{FHA22,DNS20}. Entangled light-electron states are intrinsically generated when a free electron couples to optical modes, such that each of the created excitations is associated with a different state of the electron, therefore involving different energies \cite{paper228} and transverse momenta \cite{paper400}. In this context, a strategy to produce distilled entanglement has been formulated based on pre-shaping the electron wave function \cite{paper400}. In addition, photon entanglement triggered by the interaction between free electrons and populated optical cavities has been proposed \cite{paper360,BRG22}.

In this work, we introduce a scheme to generate entangled polariton pairs in the guided modes of an optical waveguide using free electrons. In particular, we show that free electrons passing close and perpendicularly to a thin metal waveguide can generate counter-propagating surface-plasmon polaritons that are heralded in momentum- and energy-resolved measurements of the scattered electrons. The resulting polariton pairs are entangled in energy and direction of emission (left and right) in the waveguide. The proposed scheme capitalizes on the strong near-field enhancement associated with confined plasmons in ultrathin films to boost the light-matter interaction and excite multiple guided modes. We anticipate that the entangled plasmon pairs (and photon pairs after subsequent out-coupling) generated with this procedure under attainable experimental conditions are advantageous compared to existing photonic devices \cite{CBG23} (e.g., we estimate $\sim100$~MHz entangled pair generation rates in our calculations below). In addition, the generation process guarantees a high degree of synchronization between the two generated plasmons.

\begin{figure*}[t]
\begin{centering} \includegraphics[width=1.0\textwidth]{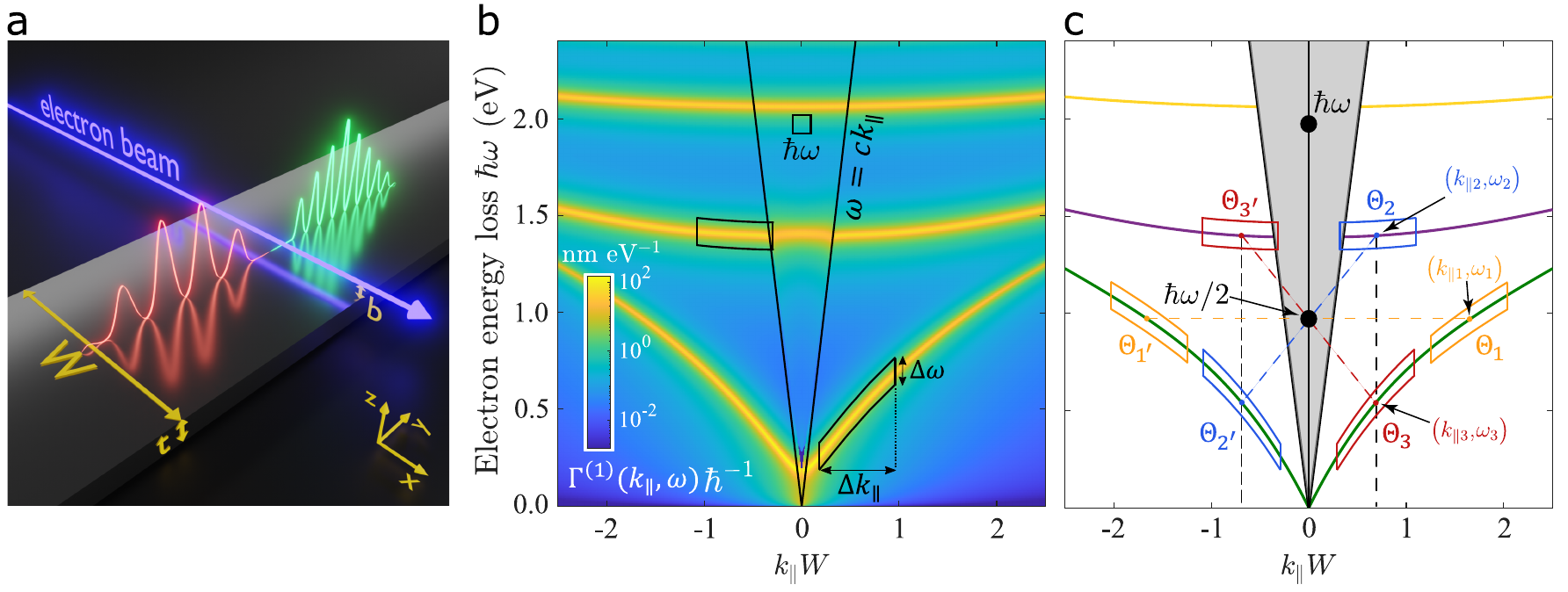} \par\end{centering}
\caption{\textbf{Generation of entangled waveguided plasmons by free electrons.} {\bf(a)}~Illustration of an electron beam that passes close and perpendicularly to a silver plasmonic waveguide (width $W$, thickness $t$) at a distance $b$ (the impact parameter) and generates two counter-propagating guided plasmons with wave vectors $k_{\parallel}$ and $-k_{\parallel}$ along the $y$ direction, respectively. {\bf(b)}~Dispersion diagram showing the wave-vector- and frequency-resolved EELS probability $\Gamma^{(1)}(\kpar,\omega)$ calculated for the silver waveguide depicted in panel (a) with parameters $W=50$~nm, $t=1.18$~nm [5 Ag(111) atomic layers], and $b=1$~nm. The electron velocity (energy) is $v=0.1\,c$ ($\approx2.6$~keV). Intense features emerge at the plasmon dispersion curves. The light cone is indicated by solid black lines, while the black square inside it shows a $(0.1/W)\times\,0.1$~eV region in $(\kpar,\hbar\omega)$ space within which electrons can be collected to herald entangled plasmon-pair emission. {\bf(c)}~ Plasmon dispersion relations in the ribbon of panel (b), where green, purple, and yellow curves correspond to the first, second, and third plasmon bands, respectively, the gray-shaded area depicts the light cone, and the regions inside the colored frames denote different combinations of entangled plasmons that can be generated when an electron loses energy $\hbar \omega$ (upper black dot) and undergoes no net momentum change along $\yy$ (undeflected electron). We define even and odd channels corresponding to the excitation of degenerate (mask pair $\{1,1'\}$) and nondegenerate (mask pairs $\{2,2'\}$ and $\{3,3'\}$) plasmon pairs. The sizes of the plasmon masks are indicated by the black contours in (b) for one of the odd-channel combinations.}
\label{Fig1}
\end{figure*}

\section*{Results}

\subsection*{Configuration for entangled polariton-pair generation}

We explore the generation of entangled waveguided polaritons in the form of propagating modes supported by optical waveguides with translation invariance in one spatial direction, taken along $\yy$. As a concrete example, we consider plasmons supported by a silver ribbon with width $W$ and thickness $t$ in the $\xx$ and $\zz$ directions, respectively, as schematically illustrated in Fig.~\ref{Fig1}a. The ribbon thicknesses are taken to span a few atomic layers, similar to those studied in recent experiments \cite{paper335}. The waveguide supports plasmon bands that are characterized by the frequency $\omega$ and wave vector $\kpar$ along $\yy$. Plasmon modes are confined to the waveguide and lie below the light cone $\omega=c\kpar$. In this configuration, a fast electron passing perpendicularly to the waveguide with velocity vector $\vb=v\,\xx$ can transfer energy $\hbar\omega$ and momentum $\hbar\kpar$ corresponding to the excitation of a plasmon mode that propagates along $\pm\yy$, but is highly confined in the transverse $x-z$ plane. The distance $b$ between the electron beam and the ribbon surface needs to be made smaller than the out-of-plane evanescent decay length of the plasmon in the surrounding vacuum (typically in the nanoscale). The frequency- and wave-vector-resolved probability that the electron creates an excitation, $\Gamma^{(1)}(\kpar,\omega)$ (Fig.~\ref{Fig1}b, calculated by following the methods in Refs.~\cite{paper040,paper149} and defined for $\omega>0$ and any real $\kpar$; see Methods) exhibits pronounced features below the light cone (solid black line) that follow the plasmon polariton dispersion relations of the waveguide and can be directly measured through EELS. The EELS dispersion diagram in Fig.~\ref{Fig1}b is obtained for an electron moving with velocity $v=0.1\,c$ ($\approx 2.6$~keV) and impact parameter $b=1$~nm above a free-standing ultrathin silver strip waveguide with dimensions $W=50$~nm and $t=1.18$~nm, corresponding to 5 Ag(111) atomic monolayers (MLs). We use the boundary-element method \cite{paper040} (BEM) with the metal response described by a background-corrected Drude model parametrized to fit the experimental data of Ref.~\cite{JC1972}: $\epsilon(\omega)=\epsilon_b-\omega_p^2/\omega(\omega+\ii\gamma)$ with $\epsilon_b=4$, $\hbar\omega_p=9.17$~eV, and $\hbar\gamma=21$~meV. We note that only a single waveguide mode can be excited at low energies, while multiple modes are available at higher energies.

Although energetic free electrons are typically assumed to undergo weak light-matter interactions that probe only the linear optical response, actual EELS measurements contain contributions due to multiple excitation events, which were observed in the first experimental evidence of the existence of surface plasmons \cite{PS1959}, while subsequent theoretical work showed that coherent states with a Poissonian population of plasmon modes are created by interaction with the free electrons \cite{SL1971,BI1983,paper228}. In particular, to second-order and neglecting quantum-coherence effects, the total EELS distribution signal is described by $\Gamma(\kpar, \omega) =  \Gamma^{(1)}(\kpar, \omega)+\Gamma^{(2)}(\kpar, \omega)$ (with units of time $\times$ distance), where the second term, quantifying the probability of exciting two light quanta (i.e., double loss events), with net energy and momentum matching those transferred from the electron, is given by the self-convolution of the linear EELS probability,
\begin{align} \label{Eq:EELS}
    &\Gamma^{(2)}(\kpar, \omega) = \\  
    &\int_{-\infty}^\infty d\kpar' \int_0^\omega d\omega'\, \Gamma^{(1)}(\kpar-\kpar', \omega-\omega')\, \Gamma^{(1)}(\kpar', \omega'). \nonumber
\end{align}
In what follows, we neglect triple and higher-order processes, which contribute negligible under the conditions here considered.

Assuming that plasmon excitations dominate the electron energy-loss signal, the leading contribution on the $\kpar=0$ axis---corresponding to electrons that lose energy but are undeflected---can be attributed to specific combinations of the guided modes excited in second-order processes, as illustrated in Fig.~\ref{Fig1}c. Detection of an electron for an energy loss $\hbar\omega$ along the $\kpar=0$ axis heralds the excitation of plasmon pairs with momentum and energy $(\hbar k_{\parallel\ell},\hbar\omega_\ell)$ and $(\hbar k_{\parallel\ell'},\hbar\omega_{\ell'})$ that satisfy $k_{\parallel\ell}=-k_{\parallel\ell'}$ and $\omega = \omega_{\ell}+\omega_{\ell'}$, where $\ell$ and $\ell'$ label plasmon pairs laying either on the same or in different plasmon bands. The geometrical construction in Fig.~\ref{Fig1}c (see dashed straight lines) shows that two different kinds of propagating-plasmon pairs can be launched, depending on whether they are either degenerate ($\hbar \omega_{\ell} =\hbar \omega_{\ell'} = \hbar \omega/2$, orange masks, which we denote as the {\it even} excitation channel) or having different energies ($\hbar \omega_{\ell} \neq \hbar \omega_{\ell'}$, red and blue masks, {\it odd} channel). For spectrally narrow plasmons, only the dots labeled $(k_{\parallel \ell},\omega_\ell)$ in the figure contribute to losses at $(\kpar=0,\omega)$. However, inelastic absorption produces plasmon broadening, which results in finite-size regions contributing to such losses, captured in our calculations through integration masks, as we discuss below.

The even channel corresponds to the simultaneous generation of two degenerate plasmons, which can be regarded as the emission of a single plasmon heralded by the detection of the other one. In contrast, the two odd-channel combinations (red and blue masks, respectively) configure an entangled plasmon pair of the form
\begin{align}
\ket{\text{left},\omega_2}\otimes\ket{\text{right},\omega_3}+\ket{\text{left},\omega_3}\otimes\ket{\text{right},\omega_2}, \label{epair}
\end{align}
formed by plasmons propagating to the left ($\kpar<0$) and to the right ($\kpar>0$) with frequencies $\omega_2\neq\omega_3$ (see Fig.~\ref{Fig1}c). We are thus interested in finding the conditions for which the odd channel dominates the total energy-loss probability of undeflected electrons (i.e., with $\kpar=0$ and $\omega=\omega_2+\omega_3$) compared even-channel plasmons as well as non-plasmonic and radiative losses.

\begin{figure*}[t!] 
\begin{centering} \includegraphics[width=1.0\textwidth]{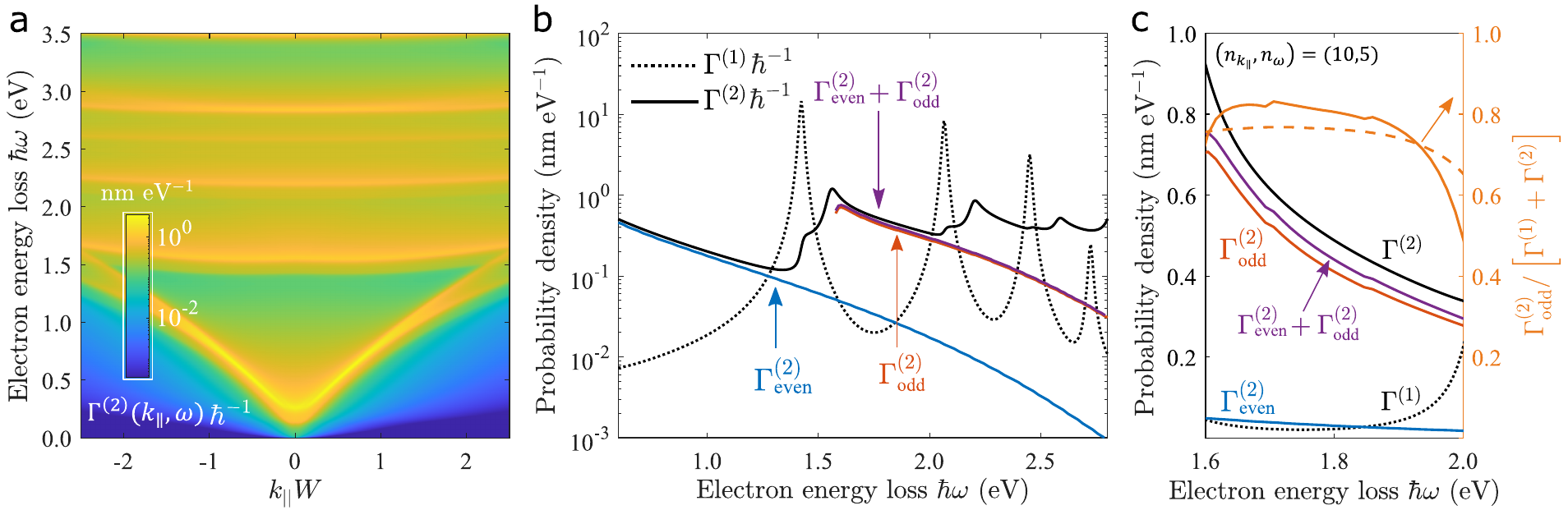} \par\end{centering}
\caption{\textbf{Quantifying entangled-plasmon generation.} {\bf(a)}~Double-loss probability obtained from the self-convolution of the EELS probability in Fig.~\ref{Fig1}b and dominated by two-plasmon events. {\bf(b)}~Calculated single- and two-plasmon excitation probabilities in a silver waveguide of width $W=10$~nm and thickness $t=0.236$~nm (1~ML) for an electron with impact parameter $b=1$~nm and velocity (energy) $v=0.02\,c$ ($\approx102$~eV) under the configuration shown in Fig.~\ref{Fig1}a. Black dotted and solid curves correspond to the total single-plasmon $\Gamma^{(1)}(\kpar=0,\omega)$ and two-plasmon $\Gamma^{(2)}(\kpar=0,\omega)$ excitation probabilities. Blue and red curves indicate the probabilities associated with even (orange masks in Fig.~\ref{Fig1}c) and odd (red and blue masks in Fig.~\ref{Fig1}c) excitation channels, respectively, while the purple curve is the sum of both contributions. {\bf(c)}~Probabilities in (b) plotted in a linear scale within a spectral region of interest (left vertical scale), along with the fraction of odd-channel events in the total two-plasmon emission probability (right scale) calculated with either the numerical BEM (solid orange curve) or the analytical 2D model (dashed orange curve).}
\label{Fig2}
\end{figure*}

\subsection*{Plasmon-pair generation with silver ribbons}

The probability that the electron undergoes two loss events is given by the self-convolution of the probability associated with single energy-loss events, as shown in Eq.~(\ref{Eq:EELS}) and represented in Fig.~\ref{Fig2}a  for a ribbon of width $W=10$~nm and thickness $t=0.236$~nm (i.e., 1~ML), considering an electron with impact parameter $b=1$~nm and velocity $v=0.02\,c$. This double-loss probability distribution features dominant contributions from two-plasmon excitations. However, other types of events such as non-plasmonic and radiative losses contribute as well (e.g., through the background signal within the light cone).

To isolate the contribution from plasmon processes that produce an EELS signal along the $\kpar=0$ axis (undeflected electrons), we specify mask functions $\Theta_\ell(\kpar,\omega)$ in momentum--frequency space, such that $\Theta_\ell$ is unity within a curved mask centered around a specific region within a given plasmon dispersion band $j$ and zero elsewhere (see Fig.~\ref{Fig1}c). More precisely, each of the curved mask areas spans a wave vector range $\pm n_{\kpar}\Delta\kpar$ around a central value $k_{\parallel \ell}$ [determined by the condition that the total wave-vector--frequency exchange is $(\kpar=0,\omega)$] as well as a frequency range $\pm n_\omega\Delta\omega$ around the $\kpar$-dependent plasmon frequency $\omega_{j\kpar}$, where $\Delta\omega=\gamma$ (the Drude damping of the metal), $\Delta\kpar=\Delta\omega/v_{j\kpar}$, $v_{j\kpar}=\partial\omega_{j\kpar}/\partial\kpar$ is the plasmon group velocity evaluated at $\kpar=k_{\parallel \ell}$, and we set $n_{\kpar}=10$ and $n_\omega=5$ in the calculations as reasonable scaling factors that allow us to have well-separated plasmons and a fair representation of the fraction of signal associated with plasmon launching. As an illustration of these definitions, the two mask contours of one of the odd-channel combinations are plotted in Fig.~\ref{Fig1}b (black).

The probababilities for even- and odd-channel plasmon-pair excitation, $\Gamma^{(2)}_{\text{even}}$ and $\Gamma^{(2)}_{\text{odd}}$, are thus calculated by restricting the convolution in Eq.~(\ref{Eq:EELS}) to the corresponding mask areas. We then have
\begin{widetext}
\begin{subequations}
\label{G2windows}
\begin{align}
&\Gamma^{(2)}_{\text{even}} (\kpar, \omega) = \int_{-\infty}^\infty d\kpar'\int_0^\omega d\omega'\,
    \Theta_{1'}(\kpar-\kpar',\omega-\omega') \Theta_1(\kpar', \omega')\; 
    \Gamma^{(1)} (\kpar- \kpar',\omega-\omega') \Gamma^{(1)}(\kpar', \omega') \label{eq:Gam_even}\\
&\Gamma^{(2)}_{\text{odd}} (\kpar, \omega) = 2\int_{-\infty}^\infty d\kpar'\int_0^\omega d\omega'\,
    \Theta_{2'}(\kpar-\kpar',\omega-\omega') \Theta_{2}(\kpar', \omega')\; 
    \Gamma^{(1)} (\kpar- \kpar',\omega-\omega') \Gamma^{(1)}(\kpar', \omega'). \label{eq:Gam_odd}
\end{align}
\end{subequations}
\end{widetext}
The factor of two in Eq.~(\ref{eq:Gam_odd}) accounts for the two possible odd-channel combinations indicated by the red and blue masks in Fig.~\ref{Fig1}c, which have the same probability (i.e., the same result is obtained when substituting the $\{2,2'\}$ subindices by $\{3,3'\}$).

In Fig.~\ref{Fig2}b, we analyze single- and two-plasmon excitation probabilities in a free-standing silver waveguide characterized by a width $W=10$~nm and a thickness $t=0.236$~nm (1~ML). We take the electron to pass with impact parameter $b=1$~nm and velocity $v=0.02\,c$ ($\approx102$~eV) above the waveguide. Specifically, we plot the probabilities for single-plasmon excitation $\Gamma^{(1)}$ (dashed black curve) and two-plasmon excitation $\Gamma^{(2)}$ (solid black curve) together with the decomposition of the latter into even [$\Gamma^{(2)}_{\text{even}}$] and odd [$\Gamma^{(2)}_{\text{odd}}$] channels. At low energies, for which the ribbon only supports a single plasmon mode, the even-channel contribution to the second-order process dominates the EELS probability. In contrast, above the single-mode threshold, $\Gamma^{(2)}_{\text{even}}$ decreases rapidly in favor of the odd-channel contribution $\Gamma^{(2)}_{\text{odd}}$ [nondegenerate entangled plasmon-pair launching, see Eq.~(\ref{epair})]. Interestingly, the single-plasmon loss channel $\Gamma^{(1)}$ displays a series of peaks associated with leaky modes inside the light cone that radiate into free space, thereby reducing the efficiency of entangled plasmon-pair production. We identify the frequency ranges for which $\Gamma^{(2)}_{\rm odd}$ is maximized and $\Gamma^{(1)}$ takes minimum values as ideal for heralding entangled plasmon-pair generation. In particular, in the $1.6-2$~eV energy-loss region, also shown in Fig.~\ref{Fig2}c in linear scale, the sum of even- and odd-channel probabilities accounts for most of the double-loss processes. Furthermore, the odd channel accounts for $\sim80\%$ of the loss probability (see Fig.~\ref{Fig2}c, orange curve, and right scale), and consequently, the detection of an electron under such circumstances signals the emission of an entangled photon pair given by Eq.~(\ref{epair}) with high probability. These results can be improved by playing with the mask parameters so that the probability associated with entangled (odd-channel) plasmon-pair emission increases with size. The contribution of even and odd channels becomes an increasingly smaller fraction of the total double-loss processes at higher loss energies due to the excitation of newly available modes, therefore degrading the fidelity of the entangled pair generation scheme. Indeed, as observed in Fig.~\ref{Fig2}a at energies above 1.5~eV, more complex combinations of higher-order modes contribute to the total EELS signal and introduce further channels to generate multiple entangled states, thus impeding the ability to confidently distinguish entangled pairs.

To quantitatively estimate the plasmon pair production efficiency, we integrate the electron scattering probability density over a region spanning $\sim0.1$\,eV in energy loss $\hbar\omega$ and $0.1/W$ in lateral wave vector $\kpar$, as indicated schematically by the small black square in Fig.\ \ref{Fig1}b. For the 10~nm silver waveguide considered in Fig.~\ref{Fig2}c, the probability density $\sim0.5$~nm~eV$^{-1}$ multiplied by the area of the considered detection window leads to $5\times10^{-4}$ entangled plasmon pairs generated per incident $\sim100$~eV electron. Considering an electron current of 1~$\mu$A ($6\times10^{12}$ electrons per second), this generation yield amounts to a rate of the order of $\sim$3\,GHz entangled pairs.

\subsection*{2D material limit}

We find the enhanced mode confinement offered by smaller waveguides produces larger yields in the free-electron generation of guided mode pairs, and thus, motivating a more exhaustive exploration of narrow 2D nanoribbons. The optical response of metallic nanoribbons with a lateral size well below the free-space light wavelengths at the frequencies of the supported plasmon resonances is accurately described in the quasistatic approximation, with the material response characterized by a 2D conductivity $\sigma(\omega) = \ii \omega t [1-\epsilon(\omega)] /4 \pi$, provided the thickness satisfies the condition $t \ll W$. (Note that we use Gaussian units throughout this work.) In this limit, we can consider a general Drude response $\sigma(\omega) = (e^2/\hbar)\,\ii\omega_D/(\omega + \ii \gamma)$ characterized by a phenomenological damping rate $\gamma$ and a weight $\omega_D$ [e.g., $\omega_D = \EF/\pi\hbar$ for graphene doped to a Fermi energy $\EF$; or also $\omega_D=\hbar\wp^2t/4\pi e^2$ for a Drude metal film of permittivity $\epsilon(\omega)=1-\wp^2/\omega(\omega+\ii\gamma)$].

Following previous works \cite{paper228,paper364}, we compute the induced field along the electron trajectory by projecting the external electron field on the eigenmodes associated with the ribbon geometry. Taking $x$ and $y$ across and along the ribbon-parallel directions (see Fig.~\ref{Fig1}a), respectively, we obtain the expression
\begin{subequations}
\label{G12D}
\begin{align}
&\Gamma^{(1)}(\kpar,\omega) = 
     \sum_j {\rm Im}\bigg\{\frac{1}{\eta_\omega^{-1}-\eta_{j\kpar}^{-1}} \bigg\}\; \mathcal{I}_j(\kpar,\omega) \label{G12Da}
\end{align}
with
\begin{align}
\mathcal{I}_j(\kpar,\omega)&=\frac{2 e^2}{\hbar v^2} \frac{\kpar^2}{\kappa^2} \ee^{-2\kappa b} \label{G12Db}\\
&\times\bigg|\int_{-W/2}^{W/2} dx\, \ee^{\ii \omega x/v}
     \vec{\EE}^*_{j\kpar}(x)\cdot\bigg(\frac{\omega}{\gamma_L^2v\kpar}\xx
     + \yy\bigg)\bigg|^2 \nonumber
\end{align}
\end{subequations}
for the EELS probability (see Methods for a self-contained derivation), where $\vec{\EE}_{j\kpar}$ are eigenmode fields labeled by the mode index $j$ and wave vector $\kpar$ that satisfy the orthonormality relation $\int_{-W/2}^{W/2} dx\; \vec{\EE}^*_{j\kpar}(x)\cdot \vec{\EE}_{j'\kpar}(x) = W\delta_{jj'}$; the corresponding real eigenvalues $\eta_{j\kpar}$ determine the plasmon resonance dispersion through the condition $\eta_\omega=\eta_{j\kpar}$ (see Methods); and the parameter $\eta_\omega = \ii \sigma(\omega)/\omega W$ encapsulates the material-dependent optical response through the 2D conductivity $\sigma(\omega)$ discussed above. Although the ribbon is treated electrostatically, relativistic effects in the interaction with the electron are incorporated through the Lorentz factor $\gamma_L =1/\sqrt{1-v^2/c^2}$, which also enters the exponential attenuation with impact parameter $b$ through $\kappa=\sqrt{\kpar^2+(\omega/v\gamma_L)^2}$. By construction, all the inelastic signal is associated with plasmon launching in this model.

We repeat the analysis in Fig.~\ref{Fig2} but use the 2D model [Eqs.~(\ref{G12D})] to calculate the loss probability. The obtained fraction of odd-channel pair generation (Fig.~\ref{Fig2}c, dashed orange curve) is similar to the BEM simulation (solid orange curve). Probabilities for first-order and second-order odd- and even-channel processes also compare reasonably well, considering that the ribbon aspect ratio $W/t\sim40$ is still not too large.

\begin{figure*}[t!] 
\begin{centering} \includegraphics[width=1.0\textwidth]{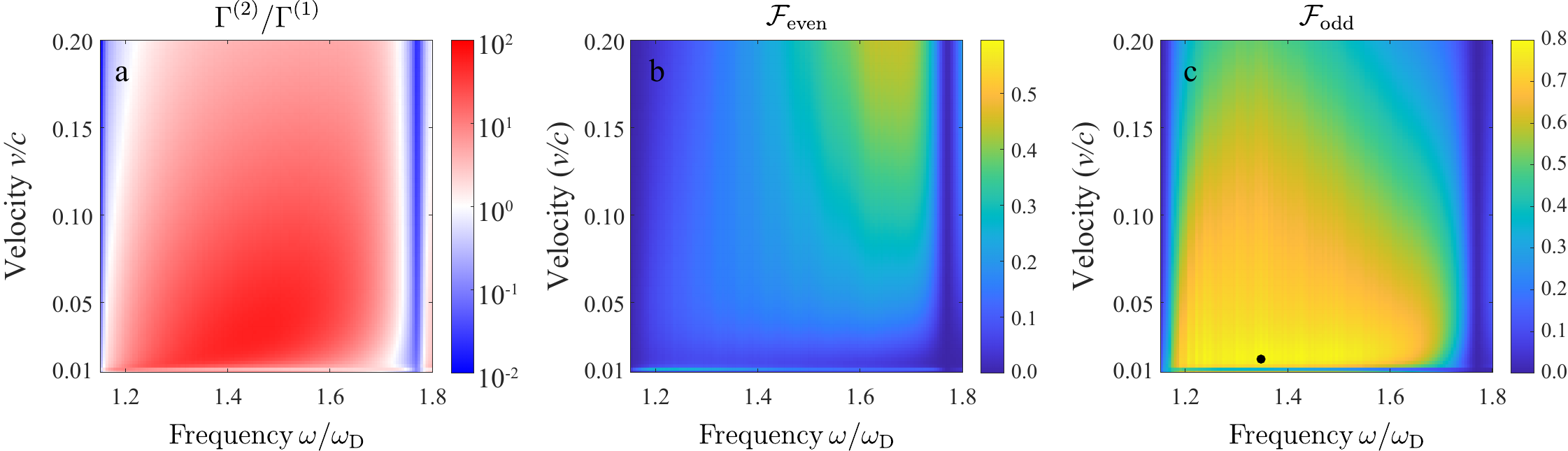} \par\end{centering}
\caption{\textbf{Universal plots of two-plasmon launching in 2D ribbons.} {\bf(a)}~Ratio of two-plasmon to single-photon excitation probabilities $\Gamma^{(2)}(0,\omega)/\Gamma^{(1)}(0,\omega)$ as a function of electron velocity $v$ and frequency loss $\omega$ (normalized to speed of light $c$ and Drude weight $\omega_D$, respectively) for undeflected electrons ($\kpar=0$) after interaction with a 2D ribbon under the conditions of Fig.~\ref{Fig1}b. The color scale is saturated at $10^{\pm2}$. {\bf(b,c)}~Fractions $\mathcal{F}_{\text{even/odd}}=\Gamma^{(2)}_{\text{even/odd}}\big/\big[\Gamma^{(1)}+\Gamma^{(2)}\big]$ (see upper labels) of undeflected electrons that generate even- and odd-channel plasmon pairs. All results are obtained from the analytical 2D model with parameters $\gamma/\omega_D=0.01$, $b/W=0.02$, and $\omega_DW/c=0.081$.}
\label{Fig3}
\end{figure*}

The analytical 2D model allows us to explore entangled plasmon-pair generation over a wide range of material, geometry, and electron parameters. In addition, upon inspection of Eqs.~(\ref{G12D}), we find that the scaled probability $cW^{-2}\Gamma^{(1)}(\kpar,\omega)$ is a function of $\kpar W$, $\omega/\omega_D$, and $v/c$ that also depends on the parameters $\omega_DW/c$, $\gamma/\omega_D$, and $b/W$. The same scaling is inherited by $cW^{-2}\Gamma^{(2)}(\kpar,\omega)$ and its even- and odd-channel contributions in virtue of Eqs.~(\ref{G2windows}). Consequently, when probabilities are integrated over wide masks (see above), the fractions of odd- and even-channel events for undeflected electrons
\begin{align}
\mathcal{F}_{\text{even/odd}}=\frac{\Gamma^{(2)}_{\text{even/odd}}(0,\omega)}{\Gamma^{(1)}(0,\omega)+\Gamma^{(2)}(0,\omega)}
\label{fractions}
\end{align}
are found to be functions of $\omega/\omega_D$, $v/c$, $\omega_DW/c$, $\gamma/\omega_D$, and $b/W$. In what follows, we set $b/W=0.02$ (grazing electron beam) and explore the dependence of $\mathcal{F}_{\rm even/odd}$ on the rest of the parameters. In addition, Fig.~\ref{Fig3}a shows the ratio of two-plasmon to single-plasmon EELS signals calculated from Eqs.~(\ref{Eq:EELS}) and (\ref{G12D}) and plotted as a function of $\omega/\omega_D$ and $v/c$ for $\gamma/\omega_D=0.01$ and $\omega_DW/c=0.081$. These parameters can be encountered in polariton-supporting ribbons made of atomically thin materials such as graphene  \cite{NMS18}, hexagonal boron nitride (hBN) \cite{GDV18}, and MoO$_3$ \cite{MAL18} films. We observe that two-plasmon events dominate over a broad range of electron velocities and energy losses. Furthermore, although the fractional contribution $\mathcal{F}_{\rm even}$ of even-channel processes is relatively small (Fig.~\ref{Fig3}b), we find that entangled (odd-channel) plasmon pairs reach a broad maximum nearing $\mathcal{F}_{\rm odd}\approx80\%$ of the total events associated with the detection of undeflected electrons (Fig.~\ref{Fig3}c). These conclusions are maintained for more lossy ribbons: the maximum fraction of entangled plasmon pairs ($\approx80\%$) is found to be rather independent of $\gamma$, while the ratio $\Gamma^{(2)}/\Gamma^{(1)}$ takes large values that scale roughly as $\propto1/\gamma$. Importantly, to obtain a large fraction of entangled plasmon-pair generation, the electron energy loss $\hbar\omega$ needs to lie between the second and third plasmon bands ($1.1\lesssim\omega/\omega_D\lesssim1.7$, approximately the range plotted in Fig.~\ref{Fig3}), while higher-order bands produce a substantial reduction in such fraction at higher energies (not shown). In addition, the absolute probability of entangled-pair generation $\Gamma^{(2)}_{\rm odd}$ exhibits a broad maximum piling up near the second plasmon band ($\omega\gtrsim1.1\omega_D$) and velocities $v\sim0.02\,c$ (not shown), where the absolute maximum is found (dot in Fig.~\ref{Fig3}c).

\section*{Discussion}

Our proposal to generate entangled photon pairs directly within optical waveguides using free electrons constitutes an alternative approach to schemes based on spontaneous parametric down-conversion of intense optical fields in bulk nonlinear optical materials. Here, the production of photons entangled in energy and momentum degrees of freedom can be heralded by the detection of electrons that lose a given amount of energy and are undeflected in the waveguide direction. To explore the feasibility of the present scheme, we have simulated the probabilities of generating counter-propagating plasmons supported by ultrathin silver waveguides that facilitate strong light-matter interactions with electrons passing along a grazing perpendicular trajectory at nonrelativistic velocities that can be achieved in integrated electron sources. Our results indicate that the enhanced field confinement offered by thin (a few atomic layers) plasmonic waveguides can improve the fidelity of entangled pair generation, which we quantify by contrasting the probability of single- and two-plasmon generation. In particular, we identify the production of nondegenerate counter-propagating plasmons (odd excitation channel) at low frequencies as the optimal configuration to herald entangled pairs with minor contamination from undesired single-plasmon emission and non-guided radiated photons.

In a practical example using silver ribbons, we predict that $\sim10^{-4}-10^{-5}$ plasmon pairs are generated per incident electron under attainable conditions by bombarding silver ribbons of $10-100$~nm width and few-atomic layer thickness with $10^2-10^4$~eV electrons. Considering an electron current of 1~$\mu$A, this generation yield amounts to a rate of $\sim0.1-1$~GHz entangled pairs. To avoid electron-electron Coulomb repulsion for such high currents, the beam could be spread along the length of the ribbon at the expense of a small degradation in the temporal synchronization of the two generated plasmons ($\sim1$\,fs per micron of beam width).

Our exploration of general polaritonic materials based on a rather accurate 2D model description of thin polariton-supporting ribbons shows that these results can be extended to long-lived plasmons in graphene \cite{NMS18} and hyperbolic phonon-polaritons in hBN \cite{GDV18} and MoO$_3$ \cite{MAL18}, which feature large quality factors $\omega/\gamma\sim10^2$ and should thus provide suitable platforms for the implementation of the proposed electron-driven entangled-plasmon-pair production at comparably high rates.

As the emission is heralded by the detection of undeflected electrons that have lost $\lesssim1$~eV energy (the total generated photon energy), the signal has to be separated from the background of losses associated with the direct excitation of a single quantum. We find that 80\% of all loss events of undeflected energy-filtered electrons are associated with the emission of entangled polariton pairs under a broad range of material, geometry, and electron parameters. Our proposed mechanism for free-electron-based entangled pair production, which can be straightforwardly applied to alternative nanophotonic or optical waveguide geometries of interest, constitutes a versatile and low-power approach to produce heralded photons directly within waveguiding architectures for present and future integrated quantum-photonics technologies.

\section*{Methods}

\subsection*{EELS simulations}

The classical EELS probability can be written as \cite{paper149}
\begin{subequations}
\label{G1gen}
\begin{align}
&\Gamma^{(1)}(\omega) = \frac{e}{\pi\hbar\omega} \int_{-\infty}^\infty \!\!\! dt\;
{\rm Re}\big\{\ee^{-\ii\omega t}\; \vb\cdot\Eb^{\rm ind}(\rb_0+\vb t,\omega)\big\} \label{G1gena}
\end{align}
in terms of the frequency-space self-induced electric field $\Eb^{\rm ind}(\rb_0+\vb t,\omega)$ acting back on the electron as it moves along a trajectory $\rb=\rb_0+\vb t$ with constant velocity vector $\vb$ (nonrecoil approximation \cite{paper149}) and passing by the position $\rb_0$ at time $t=0$. Given the translational invariance of the ribbon along $y$ (see Fig.~\ref{Fig1}a), we can decompose the EELS probability as $\Gamma^{(1)}(\omega)=\int_{-\infty}^\infty d\kpar\;\Gamma^{(1)}(\kpar,\omega)$, where
\begin{align}
&\Gamma^{(1)}(\kpar,\omega) = \frac{e}{2\pi^2\hbar\omega} \int_{-\infty}^\infty \!\!\! dx\;
{\rm Re}\big\{\ee^{-\ii\omega x/v} E_x^{\rm ind}(x,\kpar,b,\omega)\big\} \label{G1genb}
\end{align}
\end{subequations}
is obtained from Eq.~(\ref{G1gena}) by writing the induced field as $\Eb^{\rm ind}(\rb,\omega)=(2\pi)^{-1}\int d\kpar \,\ee^{\ii\kpar y}\,\Eb^{\rm ind}(x,\kpar,z,\omega)$ with $\vb=v\xx$. We use Eq.~(\ref{G1genb}) to calculate $\Gamma^{(1)}(\kpar,\omega)$ in Figs.~\ref{Fig1} and \ref{Fig2} using an adaptation of BEM to cope with translationally invariant geometries \cite{paper040} such as ribbons.

\subsection*{Zero-thickness 2D limit: derivation of Eqs.~(\ref{G12D})}

Figure~\ref{Fig3} is obtained from Eqs.~(\ref{G12D}), which is here derived in the limit of a zero-thickness ribbon, assuming an electrostatic response of the material through its local, frequency-dependent 2D conductivity $\sigma(\omega)$, but incorporating relativistic effects in the interaction with the electron.

We start by writing a relation for the total field $\Eb(\Rb,\omega)$ acting on each point $\Rb$ of a generic 2D structure in response to an external field $\Eb^{\rm ext}(\Rb,\omega)$ in the frequency domain. By first calculating the 2D charge density from the continuity equation as $(\ii\omega)^{-1}\nabla\cdot\jb^{\rm ind}_{\rm 2D}(\Rb,\omega)$, where $\jb^{\rm ind}_{\rm 2D}(\Rb,\omega)=\sigma(\omega)\,\Eb(\Rb,\omega)$ is the induced current density, we then obtain the induced field from the Coulomb potential, integrate by parts, and add the external field to find the self-consistent expression \cite{paper228}
\begin{align}
\Eb(\Rb,\omega)&=\Eb^{\rm ext}(\Rb,\omega) \label{Eself}\\
&+\frac{\ii\sigma(\omega)}{\omega}\int d^2\Rb'\, M(\Rb-\Rb')\cdot\Eb(\Rb',\omega), \nonumber
\end{align}
where the $2\times2$ tensor $M(\Rb)=\nabla\otimes\nabla R^{-1}$ acts on the plane of the structure, and we note that the 2D coordinates $\Rb$ and $\Rb'$ are restricted to the region occupied by the material throughout this section. The solution to Eq.~(\ref{Eself}) can be expressed in terms of a complete set of eigenmode fields $\Eb_j(\Rb)$ and real eigenvalues $\eta_j$ of $M$ defined by $\int d^2\Rb'\;M(\Rb-\Rb')\cdot\Eb_j(\Rb')=(\eta_jW)^{-1} \Eb_j(\Rb)$ \cite{paper228}, indexed by $j$, and satisfying the orthonormality relation $\int d^2\Rb\;\Eb_j^*(\Rb)\cdot\Eb_{j'}(\Rb)=\delta_{jj'}$, where $W$ is a characteristic distance of the structure (e.g., the ribbon width in the present work). This allows us to project Eq.~(\ref{Eself}) on the eigenmodes of $M$ and write an analytical solution for $\Eb(\rb,\omega)$, which, upon multiplication by the conductivity, yields the 2D induced current density
\begin{align}
\jb^{\rm ind}_{\rm 2D}(\Rb,\omega)=&\sigma(\omega)\sum_j\frac{1}{1-\eta_\omega/\eta_j}\;\Eb_j(\Rb) \nonumber\\
&\times\int d^2\Rb'\;\Eb_j^*(\Rb')\cdot\Eb^{\rm ext}(\Rb',\omega) \nonumber
\end{align}
with $\eta_\omega=\ii\sigma(\omega)/\omega W$. The induced electric field at any spatial location $\rb$ is then given by \cite{J99} $\Eb^{\rm ind}(\rb,\omega)=(\ii/\omega)\int d^2\Rb'\;\mathcal{G}_0(\rb-\Rb',\omega)\cdot\jb^{\rm ind}_{\rm 2D}(\Rb',\omega)$, where $\mathcal{G}_0(\rb,\omega)=(k^2+\nabla\otimes\nabla)\ee^{\ii kr}/r$ is the free-space electromagnetic Green tensor and $k=\omega/c$. We also write the external field as $\Eb^{\rm ext}(\rb,\omega)=(\ii/\omega)\int d^3\rb'\;\mathcal{G}_0(\rb-\rb',\omega)\cdot\jb^{\rm ext}(\rb',\omega)$ in terms of the external three-dimensional current density $\jb^{\rm ext}(\rb,t)=-e\vb\delta(\rb-\rb_0-\vb t)$ associated with the moving electron. More precisely \cite{paper149}, $\Eb^{\rm ext}(\rb,\omega)=(2e\omega/v^2\gamma_L)\ee^{\ii\omega r_\parallel/v}\big[\ii\gamma_L^{-1}K_0(\theta)\hat{\vb}-K_1(\theta)\hat{\rb}_\perp\big]$, where $\gamma_L=1/\sqrt{1-v^2/c^2}$ is the Lorentz factor, $K_m(\theta)$ are modified Bessel functions evaluated at $\theta=\omega r_\perp/v\gamma_L$, and we use the coordinates $r_\parallel=(\rb-\rb_0)\cdot\hat{\vb}$ and $\rb_\perp=\rb-\rb_0-r_\parallel\hat{\vb}$ parallel and perpendicular to the velocity $\vb$, respectively. Inserting these expressions for the external and the induced fields into Eq.~(\ref{G1gena}), and noticing, after some algebra, that the remaining time integral produces a factor $\big[\Eb^{\rm ext}(\rb,\omega)\big]^*$, we obtain the result
\begin{align}
\Gamma^{(1)}(\omega) =& \frac{W}{\pi\hbar} \sum_j
{\rm Im}\bigg\{\frac{1}{\eta_\omega^{-1}-\eta_j^{-1}}\bigg\} \label{Egeneral}\\
&\times\bigg|\int d^2\Rb'\;\Eb_j^*(\Rb')\cdot\Eb^{\rm ext}(\Rb',\omega)\bigg|^2. \nonumber
\end{align}
Equation~(\ref{Egeneral}) is general for any 2D structure whose response can be treated in the electrostatic limit, although relativistic effects are incorporated in the interaction with the electron through the use of the free-space Green tensor $\mathcal{G}_0$ and the resulting $\gamma_L$ factors.

For the ribbon in Fig.~\ref{Fig1}a, we can write the modes as $\Eb_{j\kpar}(\Rb)=\vec{\EE}_{j\kpar}(x)\,\ee^{\ii\kpar y}/\sqrt{WL}$ with the label $j$ multiplexed into $\{j,\kpar\}$, $L$ denoting the quantization length along the ribbon direction $y$, and the mode orthonormality relation now reducing to $\int_{-W/2}^{W/2}dx\;\vec{\EE}^*_{j\kpar}(x)\cdot\vec{\EE}_{j'\kpar}(x)=W\delta_{jj'}$. The sum over $j$ is also multiplexed as $\sum_j\sum_{\kpar}$ and we use the prescription $\sum_{\kpar}\rightarrow(L/2\pi)\int_{-\infty}^\infty d\kpar$. Making use of these elements, setting $\rb_0=b\zz$ and $\vb=v\xx$ (i.e., $r_\parallel=x$ and $\rb_\perp=y\yy+b\zz$), inserting the explicit expression for the external field given above, and retaining only the $\kpar$ integrand, Eq.~(\ref{Egeneral}) readily leads to
\begin{align}
&\Gamma^{(1)}(\kpar,\omega)
= \frac{2e^2\omega^2}{\pi^2\hbar\,v^4\gamma_L^2} \sum_j
{\rm Im}\bigg\{\frac{1}{\eta_\omega^{-1}-\eta_{j\kpar}^{-1}}\bigg\} \label{G1quasi}\\
&\times\bigg|\int_{-W/2}^{W/2} \!\! dx\,\ee^{\ii\omega x/v}\;\vec{\EE}^*_{j\kpar}(x)\cdot\Big[\ii\gamma_L^{-1} I_0(b)\,\xx - I_1(b)\,\yy\Big]\bigg|^2, \nonumber
\end{align}
where $I_m(b)=\!\int_{-\infty}^{\infty}\! dy\,\ee^{-\ii\kpar y}K_m\big(\omega\sqrt{b^2+y^2}/v\gamma_L\big)\,y^m/(b^2+y^2)^{m/2}$. From Eq.~(6.677-5) in Ref.~\cite{GR1980}, we find $I_0(b)=(\pi/\kappa)\ee^{-\kappa b}$ with $\kappa=\sqrt{\kpar^2+(\omega/v\gamma_L)^2}$, while $I_1(b)=(-\ii v\gamma_L\kpar/\omega)I_0(b)$ is obtained upon integration by parts. Inserting these results in Eq.~(\ref{G1quasi}) and making some factor rearrangements, we finally obtain Eqs.~(\ref{G12D}).

\section*{Acknowledgments}

We acknowledge support from ERC (Advanced Grant 789104-eNANO), the European Commission (Horizon 2020 Grants No. FET-Proactive 101017720-EBEAM and No. FET-Open 964591-SMART-electron), the Spanish MICINN (PID2020– 112625 GB-I00 and Severo Ochoa CEX2019-000910-S), the Catalan CERCA Program, and Fundaci\'os Cellex and Mir-Puig. J.~D.~C. is a Sapere Aude research leader supported by VILLUM FONDEN (grant no. 16498) and Independent Research Fund Denmark (grant no. 0165-00051B). The Center for Polariton-driven Light--Matter Interactions (POLIMA) is funded by the Danish National Research Foundation (Project No.~DNRF165).


\begin{thebibliography}{47}%
\makeatletter
\providecommand \@ifxundefined [1]{%
 \@ifx{#1\undefined}
}%
\providecommand \@ifnum [1]{%
 \ifnum #1\expandafter \@firstoftwo
 \else \expandafter \@secondoftwo
 \fi
}%
\providecommand \@ifx [1]{%
 \ifx #1\expandafter \@firstoftwo
 \else \expandafter \@secondoftwo
 \fi
}%
\providecommand \natexlab [1]{#1}%
\providecommand \enquote  [1]{``#1''}%
\providecommand \bibnamefont  [1]{#1}%
\providecommand \bibfnamefont [1]{#1}%
\providecommand \citenamefont [1]{#1}%
\providecommand \href@noop [0]{\@secondoftwo}%
\providecommand \href [0]{\begingroup \@sanitize@url \@href}%
\providecommand \@href[1]{\@@startlink{#1}\@@href}%
\providecommand \@@href[1]{\endgroup#1\@@endlink}%
\providecommand \@sanitize@url [0]{\catcode `\\12\catcode `\$12\catcode
  `\&12\catcode `\#12\catcode `\^12\catcode `\_12\catcode `\%12\relax}%
\providecommand \@@startlink[1]{}%
\providecommand \@@endlink[0]{}%
\providecommand \url  [0]{\begingroup\@sanitize@url \@url }%
\providecommand \@url [1]{\endgroup\@href {#1}{\urlprefix }}%
\providecommand \urlprefix  [0]{URL }%
\providecommand \Eprint [0]{\href }%
\providecommand \doibase [0]{http://dx.doi.org/}%
\providecommand \selectlanguage [0]{\@gobble}%
\providecommand \bibinfo  [0]{\@secondoftwo}%
\providecommand \bibfield  [0]{\@secondoftwo}%
\providecommand \translation [1]{[#1]}%
\providecommand \BibitemOpen [0]{}%
\providecommand \bibitemStop [0]{}%
\providecommand \bibitemNoStop [0]{.\EOS\space}%
\providecommand \EOS [0]{\spacefactor3000\relax}%
\providecommand \BibitemShut  [1]{\csname bibitem#1\endcsname}%
\let\auto@bib@innerbib\@empty
\bibitem [{\citenamefont {Bennett}\ and\ \citenamefont
  {DiVincenzo}(2000)}]{BD00}%
  \BibitemOpen
  \bibfield  {author} {\bibinfo {author} {\bibfnamefont {Charles~H.}\
  \bibnamefont {Bennett}}\ and\ \bibinfo {author} {\bibfnamefont {David~P.}\
  \bibnamefont {DiVincenzo}},\ }\bibfield  {title} {\enquote {\bibinfo {title}
  {Quantum information and computation},}\ }\href {\doibase 10.1038/35005001}
  {\bibfield  {journal} {\bibinfo  {journal} {Nature}\ }\textbf {\bibinfo
  {volume} {404}},\ \bibinfo {pages} {247--255} (\bibinfo {year}
  {2000})}\BibitemShut {NoStop}%
\bibitem [{\citenamefont {Jennewein}\ \emph {et~al.}(2000)\citenamefont
  {Jennewein}, \citenamefont {Simon}, \citenamefont {Weihs}, \citenamefont
  {Weinfurter},\ and\ \citenamefont {Zeilinger}}]{JSW00}%
  \BibitemOpen
  \bibfield  {author} {\bibinfo {author} {\bibfnamefont {Thomas}\ \bibnamefont
  {Jennewein}}, \bibinfo {author} {\bibfnamefont {Christoph}\ \bibnamefont
  {Simon}}, \bibinfo {author} {\bibfnamefont {Gregor}\ \bibnamefont {Weihs}},
  \bibinfo {author} {\bibfnamefont {Harald}\ \bibnamefont {Weinfurter}}, \ and\
  \bibinfo {author} {\bibfnamefont {Anton}\ \bibnamefont {Zeilinger}},\
  }\bibfield  {title} {\enquote {\bibinfo {title} {Quantum cryptography with
  entangled photons},}\ }\href {\doibase 10.1103/PhysRevLett.84.4729}
  {\bibfield  {journal} {\bibinfo  {journal} {Phys.\ Rev.\ Lett.}\ }\textbf
  {\bibinfo {volume} {84}},\ \bibinfo {pages} {4729--4732} (\bibinfo {year}
  {2000})}\BibitemShut {NoStop}%
\bibitem [{\citenamefont {Ma}\ \emph {et~al.}(2012)\citenamefont {Ma},
  \citenamefont {Herbst}, \citenamefont {Scheidl}, \citenamefont {Wang},
  \citenamefont {Kropatschek}, \citenamefont {Naylor}, \citenamefont
  {Wittmann}, \citenamefont {Mech}, \citenamefont {Kofler}, \citenamefont
  {Anisimova}, \citenamefont {Makarov}, \citenamefont {Jennewein},
  \citenamefont {Ursin},\ and\ \citenamefont {Zeilinger}}]{MHS12}%
  \BibitemOpen
  \bibfield  {author} {\bibinfo {author} {\bibfnamefont {Xiao-Song}\
  \bibnamefont {Ma}}, \bibinfo {author} {\bibfnamefont {Thomas}\ \bibnamefont
  {Herbst}}, \bibinfo {author} {\bibfnamefont {Thomas}\ \bibnamefont
  {Scheidl}}, \bibinfo {author} {\bibfnamefont {Daqing}\ \bibnamefont {Wang}},
  \bibinfo {author} {\bibfnamefont {Sebastian}\ \bibnamefont {Kropatschek}},
  \bibinfo {author} {\bibfnamefont {William}\ \bibnamefont {Naylor}}, \bibinfo
  {author} {\bibfnamefont {Bernhard}\ \bibnamefont {Wittmann}}, \bibinfo
  {author} {\bibfnamefont {Alexandra}\ \bibnamefont {Mech}}, \bibinfo {author}
  {\bibfnamefont {Johannes}\ \bibnamefont {Kofler}}, \bibinfo {author}
  {\bibfnamefont {Elena}\ \bibnamefont {Anisimova}}, \bibinfo {author}
  {\bibfnamefont {Vadim}\ \bibnamefont {Makarov}}, \bibinfo {author}
  {\bibfnamefont {Thomas}\ \bibnamefont {Jennewein}}, \bibinfo {author}
  {\bibfnamefont {Rupert}\ \bibnamefont {Ursin}}, \ and\ \bibinfo {author}
  {\bibfnamefont {Anton}\ \bibnamefont {Zeilinger}},\ }\bibfield  {title}
  {\enquote {\bibinfo {title} {Quantum teleportation over 143 kilometres using
  active feed-forward},}\ }\href {\doibase 10.1038/nature11472} {\bibfield
  {journal} {\bibinfo  {journal} {Nature}\ }\textbf {\bibinfo {volume} {489}},\
  \bibinfo {pages} {269--273} (\bibinfo {year} {2012})}\BibitemShut {NoStop}%
\bibitem [{\citenamefont {Mair}\ \emph {et~al.}(2001)\citenamefont {Mair},
  \citenamefont {Vaziri}, \citenamefont {Weihs},\ and\ \citenamefont
  {Zeilinger}}]{MVW01}%
  \BibitemOpen
  \bibfield  {author} {\bibinfo {author} {\bibfnamefont {Alois}\ \bibnamefont
  {Mair}}, \bibinfo {author} {\bibfnamefont {Alipasha}\ \bibnamefont {Vaziri}},
  \bibinfo {author} {\bibfnamefont {Gregor}\ \bibnamefont {Weihs}}, \ and\
  \bibinfo {author} {\bibfnamefont {Anton}\ \bibnamefont {Zeilinger}},\
  }\bibfield  {title} {\enquote {\bibinfo {title} {Entanglement of the orbital
  angular momentum states of photons},}\ }\href {\doibase 10.1038/35085529}
  {\bibfield  {journal} {\bibinfo  {journal} {Nature}\ }\textbf {\bibinfo
  {volume} {412}},\ \bibinfo {pages} {313--316} (\bibinfo {year}
  {2001})}\BibitemShut {NoStop}%
\bibitem [{\citenamefont {Fickler}\ \emph {et~al.}(2012)\citenamefont
  {Fickler}, \citenamefont {Lapkiewicz}, \citenamefont {Plick}, \citenamefont
  {Krenn}, \citenamefont {Schaeff}, \citenamefont {Ramelow},\ and\
  \citenamefont {Zeilinger}}]{FLP12}%
  \BibitemOpen
  \bibfield  {author} {\bibinfo {author} {\bibfnamefont {Robert}\ \bibnamefont
  {Fickler}}, \bibinfo {author} {\bibfnamefont {Radek}\ \bibnamefont
  {Lapkiewicz}}, \bibinfo {author} {\bibfnamefont {William~N}\ \bibnamefont
  {Plick}}, \bibinfo {author} {\bibfnamefont {Mario}\ \bibnamefont {Krenn}},
  \bibinfo {author} {\bibfnamefont {Christoph}\ \bibnamefont {Schaeff}},
  \bibinfo {author} {\bibfnamefont {Sven}\ \bibnamefont {Ramelow}}, \ and\
  \bibinfo {author} {\bibfnamefont {Anton}\ \bibnamefont {Zeilinger}},\
  }\bibfield  {title} {\enquote {\bibinfo {title} {Quantum entanglement of high
  angular momenta},}\ }\href {\doibase 10.1126/science.1227193} {\bibfield
  {journal} {\bibinfo  {journal} {Science}\ }\textbf {\bibinfo {volume}
  {338}},\ \bibinfo {pages} {640--643} (\bibinfo {year} {2012})}\BibitemShut
  {NoStop}%
\bibitem [{\citenamefont {Krutyanskiy}\ \emph {et~al.}(2023)\citenamefont
  {Krutyanskiy}, \citenamefont {Galli}, \citenamefont {Krcmarsky},
  \citenamefont {Baier}, \citenamefont {Fioretto}, \citenamefont {Pu},
  \citenamefont {Mazloom}, \citenamefont {Sekatski}, \citenamefont {Canteri},
  \citenamefont {Teller}, \citenamefont {Schupp}, \citenamefont {Bate},
  \citenamefont {Meraner}, \citenamefont {Sangouard}, \citenamefont {Lanyon},\
  and\ \citenamefont {Northup}}]{KGK23}%
  \BibitemOpen
  \bibfield  {author} {\bibinfo {author} {\bibfnamefont {V.}~\bibnamefont
  {Krutyanskiy}}, \bibinfo {author} {\bibfnamefont {M.}~\bibnamefont {Galli}},
  \bibinfo {author} {\bibfnamefont {V.}~\bibnamefont {Krcmarsky}}, \bibinfo
  {author} {\bibfnamefont {S.}~\bibnamefont {Baier}}, \bibinfo {author}
  {\bibfnamefont {D.~A.}\ \bibnamefont {Fioretto}}, \bibinfo {author}
  {\bibfnamefont {Y.}~\bibnamefont {Pu}}, \bibinfo {author} {\bibfnamefont
  {A.}~\bibnamefont {Mazloom}}, \bibinfo {author} {\bibfnamefont
  {P.}~\bibnamefont {Sekatski}}, \bibinfo {author} {\bibfnamefont
  {M.}~\bibnamefont {Canteri}}, \bibinfo {author} {\bibfnamefont
  {M.}~\bibnamefont {Teller}}, \bibinfo {author} {\bibfnamefont
  {J.}~\bibnamefont {Schupp}}, \bibinfo {author} {\bibfnamefont
  {J.}~\bibnamefont {Bate}}, \bibinfo {author} {\bibfnamefont {M.}~\bibnamefont
  {Meraner}}, \bibinfo {author} {\bibfnamefont {N.}~\bibnamefont {Sangouard}},
  \bibinfo {author} {\bibfnamefont {B.~P.}\ \bibnamefont {Lanyon}}, \ and\
  \bibinfo {author} {\bibfnamefont {T.~E.}\ \bibnamefont {Northup}},\
  }\bibfield  {title} {\enquote {\bibinfo {title} {Entanglement of trapped-ion
  qubits separated by 230 meters},}\ }\href {\doibase
  10.1103/PhysRevLett.130.050803} {\bibfield  {journal} {\bibinfo  {journal}
  {Phys.\ Rev.\ Lett.}\ }\textbf {\bibinfo {volume} {130}},\ \bibinfo {pages}
  {050803} (\bibinfo {year} {2023})}\BibitemShut {NoStop}%
\bibitem [{\citenamefont {Kwiat}\ \emph {et~al.}(1995)\citenamefont {Kwiat},
  \citenamefont {Mattle}, \citenamefont {Weinfurter}, \citenamefont
  {Zeilinger}, \citenamefont {Sergienko},\ and\ \citenamefont {Shih}}]{KMW95}%
  \BibitemOpen
  \bibfield  {author} {\bibinfo {author} {\bibfnamefont {Paul~G}\ \bibnamefont
  {Kwiat}}, \bibinfo {author} {\bibfnamefont {Klaus}\ \bibnamefont {Mattle}},
  \bibinfo {author} {\bibfnamefont {Harald}\ \bibnamefont {Weinfurter}},
  \bibinfo {author} {\bibfnamefont {Anton}\ \bibnamefont {Zeilinger}}, \bibinfo
  {author} {\bibfnamefont {Alexander~V}\ \bibnamefont {Sergienko}}, \ and\
  \bibinfo {author} {\bibfnamefont {Yanhua}\ \bibnamefont {Shih}},\ }\bibfield
  {title} {\enquote {\bibinfo {title} {New high-intensity source of
  polarization-entangled photon pairs},}\ }\href {\doibase
  10.1103/PhysRevLett.75.4337} {\bibfield  {journal} {\bibinfo  {journal}
  {Phys.\ Rev.\ Lett.}\ }\textbf {\bibinfo {volume} {75}},\ \bibinfo {pages}
  {4337--4341} (\bibinfo {year} {1995})}\BibitemShut {NoStop}%
\bibitem [{\citenamefont {Kwiat}\ \emph {et~al.}(1999)\citenamefont {Kwiat},
  \citenamefont {Waks}, \citenamefont {White}, \citenamefont {Appelbaum},\ and\
  \citenamefont {Eberhard}}]{KWW99}%
  \BibitemOpen
  \bibfield  {author} {\bibinfo {author} {\bibfnamefont {Paul~G}\ \bibnamefont
  {Kwiat}}, \bibinfo {author} {\bibfnamefont {Edo}\ \bibnamefont {Waks}},
  \bibinfo {author} {\bibfnamefont {Andrew~G}\ \bibnamefont {White}}, \bibinfo
  {author} {\bibfnamefont {Ian}\ \bibnamefont {Appelbaum}}, \ and\ \bibinfo
  {author} {\bibfnamefont {Philippe~H}\ \bibnamefont {Eberhard}},\ }\bibfield
  {title} {\enquote {\bibinfo {title} {Ultrabright source of
  polarization-entangled photons},}\ }\href {\doibase 10.1103/PhysRevA.60.R773}
  {\bibfield  {journal} {\bibinfo  {journal} {Phys.\ Rev.\ A}\ }\textbf
  {\bibinfo {volume} {60}},\ \bibinfo {pages} {R773--R776} (\bibinfo {year}
  {1999})}\BibitemShut {NoStop}%
\bibitem [{\citenamefont {Arnaut}\ and\ \citenamefont {Barbosa}(2000)}]{AB00}%
  \BibitemOpen
  \bibfield  {author} {\bibinfo {author} {\bibfnamefont {H.~H.}\ \bibnamefont
  {Arnaut}}\ and\ \bibinfo {author} {\bibfnamefont {G.~A.}\ \bibnamefont
  {Barbosa}},\ }\bibfield  {title} {\enquote {\bibinfo {title} {Orbital and
  intrinsic angular momentum of single photons and entangled pairs of photons
  generated by parametric down-conversion},}\ }\href {\doibase
  10.1103/PhysRevLett.85.286} {\bibfield  {journal} {\bibinfo  {journal}
  {Phys.\ Rev.\ Lett.}\ }\textbf {\bibinfo {volume} {85}},\ \bibinfo {pages}
  {286--289} (\bibinfo {year} {2000})}\BibitemShut {NoStop}%
\bibitem [{\citenamefont {Boyd}(2020)}]{B20_2}%
  \BibitemOpen
  \bibfield  {author} {\bibinfo {author} {\bibfnamefont {Robert~W.}\
  \bibnamefont {Boyd}},\ }\href@noop {} {\emph {\bibinfo {title} {Nonlinear
  Optics}}}\ (\bibinfo  {publisher} {Academic press},\ \bibinfo {address}
  {Amsterdam},\ \bibinfo {year} {2020})\BibitemShut {NoStop}%
\bibitem [{\citenamefont {Ilchenko}\ \emph {et~al.}(2004)\citenamefont
  {Ilchenko}, \citenamefont {Savchenkov}, \citenamefont {Matsko},\ and\
  \citenamefont {Maleki}}]{ISM04}%
  \BibitemOpen
  \bibfield  {author} {\bibinfo {author} {\bibfnamefont {V.~S.}\ \bibnamefont
  {Ilchenko}}, \bibinfo {author} {\bibfnamefont {A.~A.}\ \bibnamefont
  {Savchenkov}}, \bibinfo {author} {\bibfnamefont {A.~B.}\ \bibnamefont
  {Matsko}}, \ and\ \bibinfo {author} {\bibfnamefont {L.}~\bibnamefont
  {Maleki}},\ }\bibfield  {title} {\enquote {\bibinfo {title} {Nonlinear optics
  and crystalline whispering gallery mode cavities},}\ }\href {\doibase
  10.1103/PhysRevLett.92.043903} {\bibfield  {journal} {\bibinfo  {journal}
  {Phys.\ Rev.\ Lett.}\ }\textbf {\bibinfo {volume} {92}},\ \bibinfo {pages}
  {043903} (\bibinfo {year} {2004})}\BibitemShut {NoStop}%
\bibitem [{\citenamefont {Yang}\ \emph {et~al.}(2008)\citenamefont {Yang},
  \citenamefont {Liscidini},\ and\ \citenamefont {Sipe}}]{YLS08}%
  \BibitemOpen
  \bibfield  {author} {\bibinfo {author} {\bibfnamefont {Zhenshan}\
  \bibnamefont {Yang}}, \bibinfo {author} {\bibfnamefont {Marco}\ \bibnamefont
  {Liscidini}}, \ and\ \bibinfo {author} {\bibfnamefont {John~E.}\ \bibnamefont
  {Sipe}},\ }\bibfield  {title} {\enquote {\bibinfo {title} {Spontaneous
  parametric down-conversion in waveguides: {A} backward heisenberg picture
  approach},}\ }\href {\doibase 10.1103/PhysRevA.77.033808} {\bibfield
  {journal} {\bibinfo  {journal} {Phys.\ Rev.\ A}\ }\textbf {\bibinfo {volume}
  {77}},\ \bibinfo {pages} {033808} (\bibinfo {year} {2008})}\BibitemShut
  {NoStop}%
\bibitem [{\citenamefont {Pomarico}\ \emph {et~al.}(2009)\citenamefont
  {Pomarico}, \citenamefont {Sanguinetti}, \citenamefont {Gisin}, \citenamefont
  {Thew}, \citenamefont {Zbinden}, \citenamefont {Schreiber}, \citenamefont
  {Thomas},\ and\ \citenamefont {Sohler}}]{PSG09}%
  \BibitemOpen
  \bibfield  {author} {\bibinfo {author} {\bibfnamefont {Enrico}\ \bibnamefont
  {Pomarico}}, \bibinfo {author} {\bibfnamefont {Bruno}\ \bibnamefont
  {Sanguinetti}}, \bibinfo {author} {\bibfnamefont {Nicolas}\ \bibnamefont
  {Gisin}}, \bibinfo {author} {\bibfnamefont {Robert}\ \bibnamefont {Thew}},
  \bibinfo {author} {\bibfnamefont {Hugo}\ \bibnamefont {Zbinden}}, \bibinfo
  {author} {\bibfnamefont {Gerhard}\ \bibnamefont {Schreiber}}, \bibinfo
  {author} {\bibfnamefont {Abu}\ \bibnamefont {Thomas}}, \ and\ \bibinfo
  {author} {\bibfnamefont {Wolfgang}\ \bibnamefont {Sohler}},\ }\bibfield
  {title} {\enquote {\bibinfo {title} {Waveguide-based {OPO} source of
  entangled photon pairs},}\ }\href {\doibase 10.1088/1367-2630/11/11/113042}
  {\bibfield  {journal} {\bibinfo  {journal} {New\ J.\ Phys.}\ }\textbf
  {\bibinfo {volume} {11}},\ \bibinfo {pages} {113042} (\bibinfo {year}
  {2009})}\BibitemShut {NoStop}%
\bibitem [{\citenamefont {Helt}\ \emph {et~al.}(2015)\citenamefont {Helt},
  \citenamefont {Steel},\ and\ \citenamefont {Sipe}}]{HSS15}%
  \BibitemOpen
  \bibfield  {author} {\bibinfo {author} {\bibfnamefont {L.~G.}\ \bibnamefont
  {Helt}}, \bibinfo {author} {\bibfnamefont {M.~J.}\ \bibnamefont {Steel}}, \
  and\ \bibinfo {author} {\bibfnamefont {J.~E.}\ \bibnamefont {Sipe}},\
  }\bibfield  {title} {\enquote {\bibinfo {title} {Spontaneous parametric
  downconversion in waveguides: what's loss got to do with it?}}\ }\href
  {\doibase 10.1088/1367-2630/17/1/013055} {\bibfield  {journal} {\bibinfo
  {journal} {New\ J.\ Phys.}\ }\textbf {\bibinfo {volume} {17}},\ \bibinfo
  {pages} {013055} (\bibinfo {year} {2015})}\BibitemShut {NoStop}%
\bibitem [{\citenamefont {Luo}\ \emph {et~al.}(2015)\citenamefont {Luo},
  \citenamefont {Herrmann}, \citenamefont {Krapick}, \citenamefont {Brecht},
  \citenamefont {Ricken}, \citenamefont {Quiring}, \citenamefont {Suche},
  \citenamefont {Sohler},\ and\ \citenamefont {Silberhorn}}]{LHK15}%
  \BibitemOpen
  \bibfield  {author} {\bibinfo {author} {\bibfnamefont {Kai-Hong}\
  \bibnamefont {Luo}}, \bibinfo {author} {\bibfnamefont {Harald}\ \bibnamefont
  {Herrmann}}, \bibinfo {author} {\bibfnamefont {Stephan}\ \bibnamefont
  {Krapick}}, \bibinfo {author} {\bibfnamefont {Benjamin}\ \bibnamefont
  {Brecht}}, \bibinfo {author} {\bibfnamefont {Raimund}\ \bibnamefont
  {Ricken}}, \bibinfo {author} {\bibfnamefont {Viktor}\ \bibnamefont
  {Quiring}}, \bibinfo {author} {\bibfnamefont {Hubertus}\ \bibnamefont
  {Suche}}, \bibinfo {author} {\bibfnamefont {Wolfgang}\ \bibnamefont
  {Sohler}}, \ and\ \bibinfo {author} {\bibfnamefont {Christine}\ \bibnamefont
  {Silberhorn}},\ }\bibfield  {title} {\enquote {\bibinfo {title} {Direct
  generation of genuine single-longitudinal-mode narrowband photon pairs},}\
  }\href {\doibase 10.1088/1367-2630/17/7/073039} {\bibfield  {journal}
  {\bibinfo  {journal} {New\ J.\ Phys.}\ }\textbf {\bibinfo {volume} {17}},\
  \bibinfo {pages} {073039} (\bibinfo {year} {2015})}\BibitemShut {NoStop}%
\bibitem [{\citenamefont {Guo}\ \emph {et~al.}(2017)\citenamefont {Guo},
  \citenamefont {ling Zou}, \citenamefont {Schuck}, \citenamefont {Jung},
  \citenamefont {Cheng},\ and\ \citenamefont {Tang}}]{GZS17}%
  \BibitemOpen
  \bibfield  {author} {\bibinfo {author} {\bibfnamefont {Xiang}\ \bibnamefont
  {Guo}}, \bibinfo {author} {\bibfnamefont {Chang}\ \bibnamefont {ling Zou}},
  \bibinfo {author} {\bibfnamefont {Carsten}\ \bibnamefont {Schuck}}, \bibinfo
  {author} {\bibfnamefont {Hojoong}\ \bibnamefont {Jung}}, \bibinfo {author}
  {\bibfnamefont {Risheng}\ \bibnamefont {Cheng}}, \ and\ \bibinfo {author}
  {\bibfnamefont {Hong~X.}\ \bibnamefont {Tang}},\ }\bibfield  {title}
  {\enquote {\bibinfo {title} {Parametric down-conversion photon-pair source on
  a nanophotonic chip},}\ }\href {\doibase 10.1038/lsa.2016.249} {\bibfield
  {journal} {\bibinfo  {journal} {Light\ Sci.\ Appl.}\ }\textbf {\bibinfo
  {volume} {6}},\ \bibinfo {pages} {e16249} (\bibinfo {year}
  {2017})}\BibitemShut {NoStop}%
\bibitem [{\citenamefont {Echarri}\ \emph {et~al.}(2022)\citenamefont
  {Echarri}, \citenamefont {Cox},\ and\ \citenamefont {{Garc\'{\i}a de
  Abajo}}}]{paper385}%
  \BibitemOpen
  \bibfield  {author} {\bibinfo {author} {\bibfnamefont {A.~Rodr{\'\i}guez}\
  \bibnamefont {Echarri}}, \bibinfo {author} {\bibfnamefont {J.~D.}\
  \bibnamefont {Cox}}, \ and\ \bibinfo {author} {\bibfnamefont {F.~J.}\
  \bibnamefont {{Garc\'{\i}a de Abajo}}},\ }\bibfield  {title} {\enquote
  {\bibinfo {title} {Direct generation of entangled photon pairs in nonlinear
  optical waveguides},}\ }\href {\doibase 10.1515/nanoph-2021-0736} {\bibfield
  {journal} {\bibinfo  {journal} {Nanophotonics}\ }\textbf {\bibinfo {volume}
  {11}},\ \bibinfo {pages} {1021--1032} (\bibinfo {year} {2022})}\BibitemShut
  {NoStop}%
\bibitem [{\citenamefont {Sun}\ \emph {et~al.}(2022)\citenamefont {Sun},
  \citenamefont {Basov},\ and\ \citenamefont {Fogler}}]{SBG22}%
  \BibitemOpen
  \bibfield  {author} {\bibinfo {author} {\bibfnamefont {Zhiyuan}\ \bibnamefont
  {Sun}}, \bibinfo {author} {\bibfnamefont {D.~N.}\ \bibnamefont {Basov}}, \
  and\ \bibinfo {author} {\bibfnamefont {M.~M.}\ \bibnamefont {Fogler}},\
  }\bibfield  {title} {\enquote {\bibinfo {title} {Graphene as a source of
  entangled plasmons},}\ }\href {\doibase 10.1103/PhysRevResearch.4.023208}
  {\bibfield  {journal} {\bibinfo  {journal} {Phys.\ Rev.\ Research}\ }\textbf
  {\bibinfo {volume} {4}},\ \bibinfo {pages} {023208} (\bibinfo {year}
  {2022})}\BibitemShut {NoStop}%
\bibitem [{\citenamefont {Tanzilli}\ \emph {et~al.}(2012)\citenamefont
  {Tanzilli}, \citenamefont {Martin}, \citenamefont {Kaiser}, \citenamefont
  {De~Micheli}, \citenamefont {Alibart},\ and\ \citenamefont
  {Ostrowsky}}]{TMK12}%
  \BibitemOpen
  \bibfield  {author} {\bibinfo {author} {\bibfnamefont {S{\'e}bastien}\
  \bibnamefont {Tanzilli}}, \bibinfo {author} {\bibfnamefont {Anthony}\
  \bibnamefont {Martin}}, \bibinfo {author} {\bibfnamefont {Florian}\
  \bibnamefont {Kaiser}}, \bibinfo {author} {\bibfnamefont {Marc~P}\
  \bibnamefont {De~Micheli}}, \bibinfo {author} {\bibfnamefont {Olivier}\
  \bibnamefont {Alibart}}, \ and\ \bibinfo {author} {\bibfnamefont {Daniel~B}\
  \bibnamefont {Ostrowsky}},\ }\bibfield  {title} {\enquote {\bibinfo {title}
  {On the genesis and evolution of integrated quantum optics},}\ }\href
  {\doibase 10.1002/lpor.201100010} {\bibfield  {journal} {\bibinfo  {journal}
  {Laser\ Photon.\ Rev.}\ }\textbf {\bibinfo {volume} {6}},\ \bibinfo {pages}
  {115--143} (\bibinfo {year} {2012})}\BibitemShut {NoStop}%
\bibitem [{\citenamefont {Fang}\ and\ \citenamefont {Sun}(2015)}]{FS15}%
  \BibitemOpen
  \bibfield  {author} {\bibinfo {author} {\bibfnamefont {Yurui}\ \bibnamefont
  {Fang}}\ and\ \bibinfo {author} {\bibfnamefont {Mengtao}\ \bibnamefont
  {Sun}},\ }\bibfield  {title} {\enquote {\bibinfo {title} {Nanoplasmonic
  waveguides: towards applications in integrated nanophotonic circuits},}\
  }\href {\doibase 10.1038/lsa.2015.67} {\bibfield  {journal} {\bibinfo
  {journal} {Light\ Sci.\ Appl.}\ }\textbf {\bibinfo {volume} {4}},\ \bibinfo
  {pages} {e294--e294} (\bibinfo {year} {2015})}\BibitemShut {NoStop}%
\bibitem [{\citenamefont {Rasmussen}\ \emph {et~al.}(2021)\citenamefont
  {Rasmussen}, \citenamefont {Gon{\c{c}}alves}, \citenamefont {Xiao},
  \citenamefont {Hofferberth}, \citenamefont {Mortensen},\ and\ \citenamefont
  {Cox}}]{RGX21}%
  \BibitemOpen
  \bibfield  {author} {\bibinfo {author} {\bibfnamefont {T.~P.}\ \bibnamefont
  {Rasmussen}}, \bibinfo {author} {\bibfnamefont {P.~A.~D.}\ \bibnamefont
  {Gon{\c{c}}alves}}, \bibinfo {author} {\bibfnamefont {Sanshui}\ \bibnamefont
  {Xiao}}, \bibinfo {author} {\bibfnamefont {Sebastian}\ \bibnamefont
  {Hofferberth}}, \bibinfo {author} {\bibfnamefont {N.~Asger}\ \bibnamefont
  {Mortensen}}, \ and\ \bibinfo {author} {\bibfnamefont {Joel~D.}\ \bibnamefont
  {Cox}},\ }\bibfield  {title} {\enquote {\bibinfo {title} {Polaritons in
  two-dimensional parabolic waveguides},}\ }\href {\doibase
  10.1021/acsphotonics.1c00481} {\bibfield  {journal} {\bibinfo  {journal}
  {ACS\ Photonics}\ }\textbf {\bibinfo {volume} {8}},\ \bibinfo {pages}
  {1840--1846} (\bibinfo {year} {2021})}\BibitemShut {NoStop}%
\bibitem [{\citenamefont {Gullans}\ \emph {et~al.}(2013)\citenamefont
  {Gullans}, \citenamefont {Chang}, \citenamefont {Koppens}, \citenamefont
  {{Garc\'{\i}a de Abajo}},\ and\ \citenamefont {Lukin}}]{paper226}%
  \BibitemOpen
  \bibfield  {author} {\bibinfo {author} {\bibfnamefont {M.}~\bibnamefont
  {Gullans}}, \bibinfo {author} {\bibfnamefont {D.~E.}\ \bibnamefont {Chang}},
  \bibinfo {author} {\bibfnamefont {F.~H.~L.}\ \bibnamefont {Koppens}},
  \bibinfo {author} {\bibfnamefont {F.~J.}\ \bibnamefont {{Garc\'{\i}a de
  Abajo}}}, \ and\ \bibinfo {author} {\bibfnamefont {M.~D.}\ \bibnamefont
  {Lukin}},\ }\bibfield  {title} {\enquote {\bibinfo {title} {Single-photon
  nonlinear optics with graphene plasmons},}\ }\href {\doibase
  10.1103/PhysRevLett.111.247401} {\bibfield  {journal} {\bibinfo  {journal}
  {Phys.\ Rev.\ Lett.}\ }\textbf {\bibinfo {volume} {111}},\ \bibinfo {pages}
  {247401} (\bibinfo {year} {2013})}\BibitemShut {NoStop}%
\bibitem [{\citenamefont {Calaj{\'o}}\ \emph {et~al.}(2023)\citenamefont
  {Calaj{\'o}}, \citenamefont {Jenke}, \citenamefont {Rozema}, \citenamefont
  {Walther}, \citenamefont {Chang},\ and\ \citenamefont {Cox}}]{CJR23}%
  \BibitemOpen
  \bibfield  {author} {\bibinfo {author} {\bibfnamefont {Giuseppe}\
  \bibnamefont {Calaj{\'o}}}, \bibinfo {author} {\bibfnamefont {Philipp~K.}\
  \bibnamefont {Jenke}}, \bibinfo {author} {\bibfnamefont {Lee~A.}\
  \bibnamefont {Rozema}}, \bibinfo {author} {\bibfnamefont {Philip}\
  \bibnamefont {Walther}}, \bibinfo {author} {\bibfnamefont {Darrick~E.}\
  \bibnamefont {Chang}}, \ and\ \bibinfo {author} {\bibfnamefont {Joel~D.}\
  \bibnamefont {Cox}},\ }\bibfield  {title} {\enquote {\bibinfo {title}
  {Nonlinear quantum logic with colliding graphene plasmons},}\ }\href
  {\doibase 10.1103/PhysRevResearch.5.013188} {\bibfield  {journal} {\bibinfo
  {journal} {Phys.\ Rev.\ Research}\ }\textbf {\bibinfo {volume} {5}},\
  \bibinfo {pages} {013188} (\bibinfo {year} {2023})}\BibitemShut {NoStop}%
\bibitem [{\citenamefont {{Garc\'{\i}a de Abajo}}(2010)}]{paper149}%
  \BibitemOpen
  \bibfield  {author} {\bibinfo {author} {\bibfnamefont {F.~J.}\ \bibnamefont
  {{Garc\'{\i}a de Abajo}}},\ }\bibfield  {title} {\enquote {\bibinfo {title}
  {Optical excitations in electron microscopy},}\ }\href {\doibase
  10.1103/RevModPhys.82.209} {\bibfield  {journal} {\bibinfo  {journal} {Rev.\
  Mod.\ Phys.}\ }\textbf {\bibinfo {volume} {82}},\ \bibinfo {pages} {209--275}
  (\bibinfo {year} {2010})}\BibitemShut {NoStop}%
\bibitem [{\citenamefont {Losquin}\ \emph {et~al.}(2015)\citenamefont
  {Losquin}, \citenamefont {Zagonel}, \citenamefont {Myroshnychenko},
  \citenamefont {Rodr\'{\i}guez-Gonz\'alez}, \citenamefont {Tenc\'e},
  \citenamefont {Scarabelli}, \citenamefont {F\"orstner}, \citenamefont
  {Liz-Marz\'an}, \citenamefont {{Garc\'{i}a de Abajo}}, \citenamefont
  {St\'ephan},\ and\ \citenamefont {Kociak}}]{paper251}%
  \BibitemOpen
  \bibfield  {author} {\bibinfo {author} {\bibfnamefont {A.}~\bibnamefont
  {Losquin}}, \bibinfo {author} {\bibfnamefont {L.~F.}\ \bibnamefont
  {Zagonel}}, \bibinfo {author} {\bibfnamefont {V.}~\bibnamefont
  {Myroshnychenko}}, \bibinfo {author} {\bibfnamefont {B.}~\bibnamefont
  {Rodr\'{\i}guez-Gonz\'alez}}, \bibinfo {author} {\bibfnamefont
  {M.}~\bibnamefont {Tenc\'e}}, \bibinfo {author} {\bibfnamefont
  {L.}~\bibnamefont {Scarabelli}}, \bibinfo {author} {\bibfnamefont
  {J.}~\bibnamefont {F\"orstner}}, \bibinfo {author} {\bibfnamefont {L.~M.}\
  \bibnamefont {Liz-Marz\'an}}, \bibinfo {author} {\bibfnamefont {F.~J.}\
  \bibnamefont {{Garc\'{i}a de Abajo}}}, \bibinfo {author} {\bibfnamefont
  {O.}~\bibnamefont {St\'ephan}}, \ and\ \bibinfo {author} {\bibfnamefont
  {M.}~\bibnamefont {Kociak}},\ }\bibfield  {title} {\enquote {\bibinfo {title}
  {Unveiling nanometer scale extinction and scattering phenomena {through}
  combined electron energy loss spectroscopy and cathodoluminescence
  measurements},}\ }\href {\doibase 10.1021/nl5043775} {\bibfield  {journal}
  {\bibinfo  {journal} {Nano\ Lett.}\ }\textbf {\bibinfo {volume} {15}},\
  \bibinfo {pages} {1229--1237} (\bibinfo {year} {2015})}\BibitemShut {NoStop}%
\bibitem [{\citenamefont {Pettit}\ \emph {et~al.}(1975)\citenamefont {Pettit},
  \citenamefont {Silcox},\ and\ \citenamefont {Vincent}}]{PSV1975}%
  \BibitemOpen
  \bibfield  {author} {\bibinfo {author} {\bibfnamefont {R.~B.}\ \bibnamefont
  {Pettit}}, \bibinfo {author} {\bibfnamefont {J.}~\bibnamefont {Silcox}}, \
  and\ \bibinfo {author} {\bibfnamefont {R.}~\bibnamefont {Vincent}},\
  }\bibfield  {title} {\enquote {\bibinfo {title} {Measurement of
  surface-plasmon dispersion in oxidized aluminum films},}\ }\href {\doibase
  10.1103/PhysRevB.11.3116} {\bibfield  {journal} {\bibinfo  {journal} {Phys.\
  Rev.\ B}\ }\textbf {\bibinfo {volume} {11}},\ \bibinfo {pages} {3116--3123}
  (\bibinfo {year} {1975})}\BibitemShut {NoStop}%
\bibitem [{\citenamefont {Egerton}(1996)}]{E96}%
  \BibitemOpen
  \bibfield  {author} {\bibinfo {author} {\bibfnamefont {Ray~F.}\ \bibnamefont
  {Egerton}},\ }\href@noop {} {\emph {\bibinfo {title} {Electron Energy-loss
  Spectroscopy in the Electron Microscope}}}\ (\bibinfo  {publisher} {Plenum
  Press},\ \bibinfo {address} {New York},\ \bibinfo {year} {1996})\BibitemShut
  {NoStop}%
\bibitem [{\citenamefont {{Benda\~{n}a}}\ \emph {et~al.}(2011)\citenamefont
  {{Benda\~{n}a}}, \citenamefont {Polman},\ and\ \citenamefont {{Garc\'{\i}a de
  Abajo}}}]{paper180}%
  \BibitemOpen
  \bibfield  {author} {\bibinfo {author} {\bibfnamefont {X.~M.}\ \bibnamefont
  {{Benda\~{n}a}}}, \bibinfo {author} {\bibfnamefont {A.}~\bibnamefont
  {Polman}}, \ and\ \bibinfo {author} {\bibfnamefont {F.~J.}\ \bibnamefont
  {{Garc\'{\i}a de Abajo}}},\ }\bibfield  {title} {\enquote {\bibinfo {title}
  {Single-photon generation by electron beams},}\ }\href {\doibase
  10.1021/nl1034732} {\bibfield  {journal} {\bibinfo  {journal} {Nano\ Lett.}\
  }\textbf {\bibinfo {volume} {11}},\ \bibinfo {pages} {5099--5103} (\bibinfo
  {year} {2011})}\BibitemShut {NoStop}%
\bibitem [{\citenamefont {Feist}\ \emph {et~al.}(2022)\citenamefont {Feist},
  \citenamefont {Huang}, \citenamefont {Arend}, \citenamefont {Yang},
  \citenamefont {Henke}, \citenamefont {Raja}, \citenamefont {Kappert},
  \citenamefont {Wang}, \citenamefont {{Louren{\c{c}}o-Martins}}, \citenamefont
  {Qiu}, \citenamefont {Liu}, \citenamefont {Kfir}, \citenamefont
  {Kippenberg},\ and\ \citenamefont {Ropers}}]{FHA22}%
  \BibitemOpen
  \bibfield  {author} {\bibinfo {author} {\bibfnamefont {A.}~\bibnamefont
  {Feist}}, \bibinfo {author} {\bibfnamefont {G.}~\bibnamefont {Huang}},
  \bibinfo {author} {\bibfnamefont {G.}~\bibnamefont {Arend}}, \bibinfo
  {author} {\bibfnamefont {Y.}~\bibnamefont {Yang}}, \bibinfo {author}
  {\bibfnamefont {J.-W.}\ \bibnamefont {Henke}}, \bibinfo {author}
  {\bibfnamefont {A.~S.}\ \bibnamefont {Raja}}, \bibinfo {author}
  {\bibfnamefont {F.~J.}\ \bibnamefont {Kappert}}, \bibinfo {author}
  {\bibfnamefont {R.~N.}\ \bibnamefont {Wang}}, \bibinfo {author}
  {\bibfnamefont {H.}~\bibnamefont {{Louren{\c{c}}o-Martins}}}, \bibinfo
  {author} {\bibfnamefont {Z.}~\bibnamefont {Qiu}}, \bibinfo {author}
  {\bibfnamefont {J.}~\bibnamefont {Liu}}, \bibinfo {author} {\bibfnamefont
  {O.}~\bibnamefont {Kfir}}, \bibinfo {author} {\bibfnamefont {T.~J.}\
  \bibnamefont {Kippenberg}}, \ and\ \bibinfo {author} {\bibfnamefont
  {C.}~\bibnamefont {Ropers}},\ }\bibfield  {title} {\enquote {\bibinfo {title}
  {Cavity-mediated electron-photon pairs},}\ }\href {\doibase
  10.1126/science.abo503} {\bibfield  {journal} {\bibinfo  {journal} {Science}\
  }\textbf {\bibinfo {volume} {377}},\ \bibinfo {pages} {777--780} (\bibinfo
  {year} {2022})}\BibitemShut {NoStop}%
\bibitem [{\citenamefont {Dahan}\ \emph {et~al.}(2020)\citenamefont {Dahan},
  \citenamefont {Nehemia}, \citenamefont {Shentcis}, \citenamefont {Reinhardt},
  \citenamefont {Adiv}, \citenamefont {Shi}, \citenamefont {Be'er},
  \citenamefont {Lynch}, \citenamefont {Kurman}, \citenamefont {Wang},\ and\
  \citenamefont {Kaminer}}]{DNS20}%
  \BibitemOpen
  \bibfield  {author} {\bibinfo {author} {\bibfnamefont {R.}~\bibnamefont
  {Dahan}}, \bibinfo {author} {\bibfnamefont {S.}~\bibnamefont {Nehemia}},
  \bibinfo {author} {\bibfnamefont {M.}~\bibnamefont {Shentcis}}, \bibinfo
  {author} {\bibfnamefont {O.}~\bibnamefont {Reinhardt}}, \bibinfo {author}
  {\bibfnamefont {Y.}~\bibnamefont {Adiv}}, \bibinfo {author} {\bibfnamefont
  {X.}~\bibnamefont {Shi}}, \bibinfo {author} {\bibfnamefont {O.}~\bibnamefont
  {Be'er}}, \bibinfo {author} {\bibfnamefont {M.~H.}\ \bibnamefont {Lynch}},
  \bibinfo {author} {\bibfnamefont {Y.}~\bibnamefont {Kurman}}, \bibinfo
  {author} {\bibfnamefont {K.}~\bibnamefont {Wang}}, \ and\ \bibinfo {author}
  {\bibfnamefont {I.}~\bibnamefont {Kaminer}},\ }\bibfield  {title} {\enquote
  {\bibinfo {title} {Resonant phase-matching {between} a light wave and a
  free-electron wavefunction},}\ }\href {\doibase 10.1038/s41567-020-01042-w}
  {\bibfield  {journal} {\bibinfo  {journal} {Nat.\ Phys.}\ }\textbf {\bibinfo
  {volume} {16}},\ \bibinfo {pages} {1123--1131} (\bibinfo {year}
  {2020})}\BibitemShut {NoStop}%
\bibitem [{\citenamefont {{Garc\'{\i}a de Abajo}}(2013)}]{paper228}%
  \BibitemOpen
  \bibfield  {author} {\bibinfo {author} {\bibfnamefont {F.~J.}\ \bibnamefont
  {{Garc\'{\i}a de Abajo}}},\ }\bibfield  {title} {\enquote {\bibinfo {title}
  {Multiple excitation of confined graphene plasmons by single free
  electrons},}\ }\href {\doibase 10.1021/nn405367e} {\bibfield  {journal}
  {\bibinfo  {journal} {ACS\ Nano}\ }\textbf {\bibinfo {volume} {7}},\ \bibinfo
  {pages} {11409--11419} (\bibinfo {year} {2013})}\BibitemShut {NoStop}%
\bibitem [{\citenamefont {Kone\v{c}n\'{a}}\ \emph {et~al.}(2022)\citenamefont
  {Kone\v{c}n\'{a}}, \citenamefont {Iyikanat},\ and\ \citenamefont
  {{Garc\'{\i}a de Abajo}}}]{paper400}%
  \BibitemOpen
  \bibfield  {author} {\bibinfo {author} {\bibfnamefont {A.}~\bibnamefont
  {Kone\v{c}n\'{a}}}, \bibinfo {author} {\bibfnamefont {F.}~\bibnamefont
  {Iyikanat}}, \ and\ \bibinfo {author} {\bibfnamefont {F.~J.}\ \bibnamefont
  {{Garc\'{\i}a de Abajo}}},\ }\bibfield  {title} {\enquote {\bibinfo {title}
  {Entangling free electrons and optical excitations},}\ }\href {\doibase
  10.1126/sciadv.abo7853} {\bibfield  {journal} {\bibinfo  {journal} {Sci.\
  Adv.}\ }\textbf {\bibinfo {volume} {8}},\ \bibinfo {pages} {eabo7853}
  (\bibinfo {year} {2022})}\BibitemShut {NoStop}%
\bibitem [{\citenamefont {{Di Giulio}}\ and\ \citenamefont {{Garc\'{\i}a de
  Abajo}}(2020)}]{paper360}%
  \BibitemOpen
  \bibfield  {author} {\bibinfo {author} {\bibfnamefont {V.}~\bibnamefont {{Di
  Giulio}}}\ and\ \bibinfo {author} {\bibfnamefont {F.~J.}\ \bibnamefont
  {{Garc\'{\i}a de Abajo}}},\ }\bibfield  {title} {\enquote {\bibinfo {title}
  {Free-electron shaping using quantum light},}\ }\href {\doibase
  10.1364/OPTICA.404598} {\bibfield  {journal} {\bibinfo  {journal} {Optica}\
  }\textbf {\bibinfo {volume} {7}},\ \bibinfo {pages} {1820--1830} (\bibinfo
  {year} {2020})}\BibitemShut {NoStop}%
\bibitem [{\citenamefont {Baranes}\ \emph {et~al.}(2022)\citenamefont
  {Baranes}, \citenamefont {Ruimy}, \citenamefont {Gorlach},\ and\
  \citenamefont {Kaminer}}]{BRG22}%
  \BibitemOpen
  \bibfield  {author} {\bibinfo {author} {\bibfnamefont {Gefen}\ \bibnamefont
  {Baranes}}, \bibinfo {author} {\bibfnamefont {Ron}\ \bibnamefont {Ruimy}},
  \bibinfo {author} {\bibfnamefont {Alexey}\ \bibnamefont {Gorlach}}, \ and\
  \bibinfo {author} {\bibfnamefont {Ido}\ \bibnamefont {Kaminer}},\ }\bibfield
  {title} {\enquote {\bibinfo {title} {Free electrons can induce entanglement
  between photons},}\ }\href {\doibase 10.1038/s41534-022-00540-4} {\bibfield
  {journal} {\bibinfo  {journal} {npj\ Quantum\ Inf.}\ }\textbf {\bibinfo
  {volume} {8}},\ \bibinfo {pages} {32} (\bibinfo {year} {2022})}\BibitemShut
  {NoStop}%
\bibitem [{\citenamefont {Chopin}\ \emph {et~al.}(2023)\citenamefont {Chopin},
  \citenamefont {Barone}, \citenamefont {Ghorbel}, \citenamefont {Combri\'e},
  \citenamefont {Bajoni}, \citenamefont {Raineri}, \citenamefont {Galli},\ and\
  \citenamefont {{De Rossi}}}]{CBG23}%
  \BibitemOpen
  \bibfield  {author} {\bibinfo {author} {\bibfnamefont {A.}~\bibnamefont
  {Chopin}}, \bibinfo {author} {\bibfnamefont {A.}~\bibnamefont {Barone}},
  \bibinfo {author} {\bibfnamefont {I.}~\bibnamefont {Ghorbel}}, \bibinfo
  {author} {\bibfnamefont {S.}~\bibnamefont {Combri\'e}}, \bibinfo {author}
  {\bibfnamefont {D.}~\bibnamefont {Bajoni}}, \bibinfo {author} {\bibfnamefont
  {F.}~\bibnamefont {Raineri}}, \bibinfo {author} {\bibfnamefont
  {M.}~\bibnamefont {Galli}}, \ and\ \bibinfo {author} {\bibfnamefont
  {A.}~\bibnamefont {{De Rossi}}},\ }\bibfield  {title} {\enquote {\bibinfo
  {title} {Ultra-efficient generation of time-energy entangled photon pairs in
  an {InGaP} photonic crystal cavity},}\ }\href {\doibase
  10.1038/s42005-023-01189-x} {\bibfield  {journal} {\bibinfo  {journal}
  {Commun.\ Phys.}\ }\textbf {\bibinfo {volume} {6}},\ \bibinfo {pages} {77}
  (\bibinfo {year} {2023})}\BibitemShut {NoStop}%
\bibitem [{\citenamefont {{Abd El-Fattah}}\ \emph {et~al.}(2019)\citenamefont
  {{Abd El-Fattah}}, \citenamefont {Mkhitaryan}, \citenamefont {Brede},
  \citenamefont {Fern\'andez}, \citenamefont {Li}, \citenamefont {Guo},
  \citenamefont {Ghosh}, \citenamefont {{Rodr\'{\i}guez Echarri}},
  \citenamefont {Naveh}, \citenamefont {Xia}, \citenamefont {Ortega},\ and\
  \citenamefont {{Garc\'{\i}a de Abajo}}}]{paper335}%
  \BibitemOpen
  \bibfield  {author} {\bibinfo {author} {\bibfnamefont {Z.~M.}\ \bibnamefont
  {{Abd El-Fattah}}}, \bibinfo {author} {\bibfnamefont {V.}~\bibnamefont
  {Mkhitaryan}}, \bibinfo {author} {\bibfnamefont {J.}~\bibnamefont {Brede}},
  \bibinfo {author} {\bibfnamefont {L.}~\bibnamefont {Fern\'andez}}, \bibinfo
  {author} {\bibfnamefont {C.}~\bibnamefont {Li}}, \bibinfo {author}
  {\bibfnamefont {Q.}~\bibnamefont {Guo}}, \bibinfo {author} {\bibfnamefont
  {A.}~\bibnamefont {Ghosh}}, \bibinfo {author} {\bibfnamefont
  {A.}~\bibnamefont {{Rodr\'{\i}guez Echarri}}}, \bibinfo {author}
  {\bibfnamefont {D.}~\bibnamefont {Naveh}}, \bibinfo {author} {\bibfnamefont
  {F.}~\bibnamefont {Xia}}, \bibinfo {author} {\bibfnamefont {J.~E.}\
  \bibnamefont {Ortega}}, \ and\ \bibinfo {author} {\bibfnamefont {F.~J.}\
  \bibnamefont {{Garc\'{\i}a de Abajo}}},\ }\bibfield  {title} {\enquote
  {\bibinfo {title} {Plasmonics in atomically thin crystalline silver films},}\
  }\href {\doibase 10.1021/acsnano.9b01651} {\bibfield  {journal} {\bibinfo
  {journal} {ACS\ Nano}\ }\textbf {\bibinfo {volume} {13}},\ \bibinfo {pages}
  {7771--7779} (\bibinfo {year} {2019})}\BibitemShut {NoStop}%
\bibitem [{\citenamefont {{Garc\'{\i}a de Abajo}}\ and\ \citenamefont
  {Howie}(2002)}]{paper040}%
  \BibitemOpen
  \bibfield  {author} {\bibinfo {author} {\bibfnamefont {F.~J.}\ \bibnamefont
  {{Garc\'{\i}a de Abajo}}}\ and\ \bibinfo {author} {\bibfnamefont
  {A.}~\bibnamefont {Howie}},\ }\bibfield  {title} {\enquote {\bibinfo {title}
  {Retarded field calculation of electron energy loss in inhomogeneous
  dielectrics},}\ }\href {\doibase 10.1103/PhysRevB.65.115418} {\bibfield
  {journal} {\bibinfo  {journal} {Phys.\ Rev.\ B}\ }\textbf {\bibinfo {volume}
  {65}},\ \bibinfo {pages} {115418} (\bibinfo {year} {2002})}\BibitemShut
  {NoStop}%
\bibitem [{\citenamefont {Johnson}\ and\ \citenamefont
  {Christy}(1972)}]{JC1972}%
  \BibitemOpen
  \bibfield  {author} {\bibinfo {author} {\bibfnamefont {P.~B.}\ \bibnamefont
  {Johnson}}\ and\ \bibinfo {author} {\bibfnamefont {R.~W.}\ \bibnamefont
  {Christy}},\ }\bibfield  {title} {\enquote {\bibinfo {title} {Optical
  constants of the noble metals},}\ }\href {\doibase 10.1103/PhysRevB.6.4370}
  {\bibfield  {journal} {\bibinfo  {journal} {Phys.\ Rev.\ B}\ }\textbf
  {\bibinfo {volume} {6}},\ \bibinfo {pages} {4370--4379} (\bibinfo {year}
  {1972})}\BibitemShut {NoStop}%
\bibitem [{\citenamefont {Powell}\ and\ \citenamefont {Swan}(1959)}]{PS1959}%
  \BibitemOpen
  \bibfield  {author} {\bibinfo {author} {\bibfnamefont {C.~J.}\ \bibnamefont
  {Powell}}\ and\ \bibinfo {author} {\bibfnamefont {J.~B.}\ \bibnamefont
  {Swan}},\ }\bibfield  {title} {\enquote {\bibinfo {title} {Origin of the
  characteristic electron energy losses in aluminum},}\ }\href {\doibase
  10.1103/PhysRev.115.869} {\bibfield  {journal} {\bibinfo  {journal} {Phys.\
  Rev.}\ }\textbf {\bibinfo {volume} {115}},\ \bibinfo {pages} {869--875}
  (\bibinfo {year} {1959})}\BibitemShut {NoStop}%
\bibitem [{\citenamefont {{\v{S}unji\'{c}}}\ and\ \citenamefont
  {Lucas}(1971)}]{SL1971}%
  \BibitemOpen
  \bibfield  {author} {\bibinfo {author} {\bibfnamefont {M.}~\bibnamefont
  {{\v{S}unji\'{c}}}}\ and\ \bibinfo {author} {\bibfnamefont {A.~A.}\
  \bibnamefont {Lucas}},\ }\bibfield  {title} {\enquote {\bibinfo {title}
  {Multiple plasmon effects in the energy-loss spectra of electrons in thin
  films},}\ }\href {\doibase 10.1103/PhysRevB.3.719} {\bibfield  {journal}
  {\bibinfo  {journal} {Phys.\ Rev.\ B}\ }\textbf {\bibinfo {volume} {3}},\
  \bibinfo {pages} {719--729} (\bibinfo {year} {1971})}\BibitemShut {NoStop}%
\bibitem [{\citenamefont {Backes}\ and\ \citenamefont {Ibach}(1983)}]{BI1983}%
  \BibitemOpen
  \bibfield  {author} {\bibinfo {author} {\bibfnamefont {U}~\bibnamefont
  {Backes}}\ and\ \bibinfo {author} {\bibfnamefont {H}~\bibnamefont {Ibach}},\
  }\bibfield  {title} {\enquote {\bibinfo {title} {Electron energy losses from
  thin silver films},}\ }\href {\doibase 10.1016/0038-1098(83)90850-5}
  {\bibfield  {journal} {\bibinfo  {journal} {Solid\ State\ Commun.}\ }\textbf
  {\bibinfo {volume} {48}},\ \bibinfo {pages} {445--447} (\bibinfo {year}
  {1983})}\BibitemShut {NoStop}%
\bibitem [{\citenamefont {Mkhitaryan}\ \emph {et~al.}(2021)\citenamefont
  {Mkhitaryan}, \citenamefont {Dias}, \citenamefont {Carbone},\ and\
  \citenamefont {{Garc\'{\i}a de Abajo}}}]{paper364}%
  \BibitemOpen
  \bibfield  {author} {\bibinfo {author} {\bibfnamefont {V.}~\bibnamefont
  {Mkhitaryan}}, \bibinfo {author} {\bibfnamefont {E.~J.~C.}\ \bibnamefont
  {Dias}}, \bibinfo {author} {\bibfnamefont {F.}~\bibnamefont {Carbone}}, \
  and\ \bibinfo {author} {\bibfnamefont {F.~J.}\ \bibnamefont {{Garc\'{\i}a de
  Abajo}}},\ }\bibfield  {title} {\enquote {\bibinfo {title} {Ultrafast
  momentum-resolved free-electron probing of optically pumped plasmon thermal
  dynamics},}\ }\href {\doibase 10.1021/acsphotonics.0c01758} {\bibfield
  {journal} {\bibinfo  {journal} {ACS\ Photonics}\ }\textbf {\bibinfo {volume}
  {8}},\ \bibinfo {pages} {614--624} (\bibinfo {year} {2021})}\BibitemShut
  {NoStop}%
\bibitem [{\citenamefont {{Ni}}\ \emph {et~al.}(2018)\citenamefont {{Ni}},
  \citenamefont {McLeod}, \citenamefont {Sun}, \citenamefont {Wang},
  \citenamefont {Xiong}, \citenamefont {Post}, \citenamefont {Sunku},
  \citenamefont {Jiang}, \citenamefont {Hone}, \citenamefont {Dean},
  \citenamefont {Fogler},\ and\ \citenamefont {Basov}}]{NMS18}%
  \BibitemOpen
  \bibfield  {author} {\bibinfo {author} {\bibfnamefont {G.~X.}\ \bibnamefont
  {{Ni}}}, \bibinfo {author} {\bibfnamefont {A.~S.}\ \bibnamefont {McLeod}},
  \bibinfo {author} {\bibfnamefont {Z.}~\bibnamefont {Sun}}, \bibinfo {author}
  {\bibfnamefont {L.}~\bibnamefont {Wang}}, \bibinfo {author} {\bibfnamefont
  {L.}~\bibnamefont {Xiong}}, \bibinfo {author} {\bibfnamefont {K.~W.}\
  \bibnamefont {Post}}, \bibinfo {author} {\bibfnamefont {S.~S.}\ \bibnamefont
  {Sunku}}, \bibinfo {author} {\bibfnamefont {B.-Y.}\ \bibnamefont {Jiang}},
  \bibinfo {author} {\bibfnamefont {J.}~\bibnamefont {Hone}}, \bibinfo {author}
  {\bibfnamefont {C.~R.}\ \bibnamefont {Dean}}, \bibinfo {author}
  {\bibfnamefont {M.~M.}\ \bibnamefont {Fogler}}, \ and\ \bibinfo {author}
  {\bibfnamefont {D.~N.}\ \bibnamefont {Basov}},\ }\bibfield  {title} {\enquote
  {\bibinfo {title} {Fundamental limits to graphene plasmonics},}\ }\href
  {\doibase 10.1038/s41586-018-0136-9} {\bibfield  {journal} {\bibinfo
  {journal} {Nature}\ }\textbf {\bibinfo {volume} {557}},\ \bibinfo {pages}
  {530--533} (\bibinfo {year} {2018})}\BibitemShut {NoStop}%
\bibitem [{\citenamefont {Giles}\ \emph {et~al.}(2018)\citenamefont {Giles},
  \citenamefont {Dai}, \citenamefont {Vurgaftman}, \citenamefont {Hoffman},
  \citenamefont {Liu}, \citenamefont {Lindsay}, \citenamefont {Ellis},
  \citenamefont {Assefa}, \citenamefont {Chatzakis}, \citenamefont {Reinecke},
  \citenamefont {Tischler}, \citenamefont {Fogler}, \citenamefont {Edgar},
  \citenamefont {Basov},\ and\ \citenamefont {Caldwell}}]{GDV18}%
  \BibitemOpen
  \bibfield  {author} {\bibinfo {author} {\bibfnamefont {Alexander~J}\
  \bibnamefont {Giles}}, \bibinfo {author} {\bibfnamefont {Siyuan}\
  \bibnamefont {Dai}}, \bibinfo {author} {\bibfnamefont {Igor}\ \bibnamefont
  {Vurgaftman}}, \bibinfo {author} {\bibfnamefont {Timothy}\ \bibnamefont
  {Hoffman}}, \bibinfo {author} {\bibfnamefont {Song}\ \bibnamefont {Liu}},
  \bibinfo {author} {\bibfnamefont {Lucas}\ \bibnamefont {Lindsay}}, \bibinfo
  {author} {\bibfnamefont {Chase~T}\ \bibnamefont {Ellis}}, \bibinfo {author}
  {\bibfnamefont {Nathanael}\ \bibnamefont {Assefa}}, \bibinfo {author}
  {\bibfnamefont {Ioannis}\ \bibnamefont {Chatzakis}}, \bibinfo {author}
  {\bibfnamefont {Thomas~L}\ \bibnamefont {Reinecke}}, \bibinfo {author}
  {\bibfnamefont {Joseph~G.}\ \bibnamefont {Tischler}}, \bibinfo {author}
  {\bibfnamefont {Michael~M.}\ \bibnamefont {Fogler}}, \bibinfo {author}
  {\bibfnamefont {J.~H.}\ \bibnamefont {Edgar}}, \bibinfo {author}
  {\bibfnamefont {D.~N.}\ \bibnamefont {Basov}}, \ and\ \bibinfo {author}
  {\bibfnamefont {Joshua~D.}\ \bibnamefont {Caldwell}},\ }\bibfield  {title}
  {\enquote {\bibinfo {title} {Ultralow-loss polaritons in isotopically pure
  boron nitride},}\ }\href {\doibase 10.1038/nmat5047} {\bibfield  {journal}
  {\bibinfo  {journal} {Nat.\ Mater.}\ }\textbf {\bibinfo {volume} {17}},\
  \bibinfo {pages} {134--139} (\bibinfo {year} {2018})}\BibitemShut {NoStop}%
\bibitem [{\citenamefont {Ma1}\ \emph {et~al.}(2018)\citenamefont {Ma1},
  \citenamefont {Alonso-Gonz\'alez}, \citenamefont {Li}, \citenamefont
  {Nikitin}, \citenamefont {Yuan}, \citenamefont {Mart\i{\i}n-S\'anchez},
  \citenamefont {Taboada-Guti\'eerrez}, \citenamefont {Amenabar}, \citenamefont
  {Li}, \citenamefont {V\'elez}, \citenamefont {Tollan}, \citenamefont {Dai},
  \citenamefont {Zhang}, \citenamefont {Sriram}, \citenamefont
  {Kalantar-Zadeh}, \citenamefont {Lee}, \citenamefont {Hillenbrand},\ and\
  \citenamefont {Bao}}]{MAL18}%
  \BibitemOpen
  \bibfield  {author} {\bibinfo {author} {\bibfnamefont {Weiliang}\
  \bibnamefont {Ma1}}, \bibinfo {author} {\bibfnamefont {Pablo}\ \bibnamefont
  {Alonso-Gonz\'alez}}, \bibinfo {author} {\bibfnamefont {Shaojuan}\
  \bibnamefont {Li}}, \bibinfo {author} {\bibfnamefont {Alexey~Y.}\
  \bibnamefont {Nikitin}}, \bibinfo {author} {\bibfnamefont {Jian}\
  \bibnamefont {Yuan}}, \bibinfo {author} {\bibfnamefont {Javier}\ \bibnamefont
  {Mart\i{\i}n-S\'anchez}}, \bibinfo {author} {\bibfnamefont {Javier}\
  \bibnamefont {Taboada-Guti\'eerrez}}, \bibinfo {author} {\bibfnamefont
  {Iban}\ \bibnamefont {Amenabar}}, \bibinfo {author} {\bibfnamefont {Peining}\
  \bibnamefont {Li}}, \bibinfo {author} {\bibfnamefont {Sa\"ul}\ \bibnamefont
  {V\'elez}}, \bibinfo {author} {\bibfnamefont {Christopher}\ \bibnamefont
  {Tollan}}, \bibinfo {author} {\bibfnamefont {Zhigao}\ \bibnamefont {Dai}},
  \bibinfo {author} {\bibfnamefont {Yupeng}\ \bibnamefont {Zhang}}, \bibinfo
  {author} {\bibfnamefont {Sharath}\ \bibnamefont {Sriram}}, \bibinfo {author}
  {\bibfnamefont {Kourosh}\ \bibnamefont {Kalantar-Zadeh}}, \bibinfo {author}
  {\bibfnamefont {Shuit-Tong}\ \bibnamefont {Lee}}, \bibinfo {author}
  {\bibfnamefont {Rainer}\ \bibnamefont {Hillenbrand}}, \ and\ \bibinfo
  {author} {\bibfnamefont {Qiaoliang}\ \bibnamefont {Bao}},\ }\bibfield
  {title} {\enquote {\bibinfo {title} {In-plane anisotropic and ultra-low-loss
  polaritons in a natural van der waals crystal},}\ }\href {\doibase
  10.1038/s41586-018-0618-9} {\bibfield  {journal} {\bibinfo  {journal}
  {Nature}\ }\textbf {\bibinfo {volume} {562}},\ \bibinfo {pages} {557--562}
  (\bibinfo {year} {2018})}\BibitemShut {NoStop}%
\bibitem [{\citenamefont {Jackson}(1999)}]{J99}%
  \BibitemOpen
  \bibfield  {author} {\bibinfo {author} {\bibfnamefont {J.~D.}\ \bibnamefont
  {Jackson}},\ }\href@noop {} {\emph {\bibinfo {title} {Classical
  Electrodynamics}}}\ (\bibinfo  {publisher} {Wiley},\ \bibinfo {address} {New
  York},\ \bibinfo {year} {1999})\BibitemShut {NoStop}%
\bibitem [{\citenamefont {Gradshteyn}\ and\ \citenamefont
  {Ryzhik}(2007)}]{GR1980}%
  \BibitemOpen
  \bibfield  {author} {\bibinfo {author} {\bibfnamefont {I.~S.}\ \bibnamefont
  {Gradshteyn}}\ and\ \bibinfo {author} {\bibfnamefont {I.~M.}\ \bibnamefont
  {Ryzhik}},\ }\href@noop {} {\emph {\bibinfo {title} {Table of Integrals,
  Series, and Products}}}\ (\bibinfo  {publisher} {Academic Press},\ \bibinfo
  {address} {London},\ \bibinfo {year} {2007})\BibitemShut {NoStop}%
\end{thebibliography}

%

\end{document}